\documentclass[preprint,twocolumn]{rmaa}
\pdfoutput=1
\usepackage{rotating}
\usepackage{amsmath}
\usepackage{microtype}
\usepackage{hyperref}
\hypersetup{
  pdfauthor={Ma. T. García-Díaz, W. J. Henney, J. A. López, and T. Doi},
  pdftitle={Velocity Structure in the Orion Nebula. II. Emission Line
    Atlas of Partially Ionized to Fully Ionized Gas}
}

\setkeys{Gin}{width=\linewidth, height=0.9\textheight, keepaspectratio=true}

\newcommand{\kms}{\ensuremath{\mathrm{km\ s}^{-1}}}
\newcommand{\pcc}{\ensuremath{\mathrm{cm}^{-3}}}
\newcommand{\kmssq}{\ensuremath{\mathrm{km^2\ s}^{-2}}}
\newcommand{\thC}{$\theta^1$~Ori~C}

\newcommand{\vhel}{\ensuremath{V_\odot}}
\newcommand{\vmean}{\ensuremath{\langle V\rangle}}
\newcommand{\vpeak}{\ensuremath{V_\mathrm{p}}}
\newcommand\linelam[2]{#1 #2\,\AA\@}
\newcommand\OIlam{\linelam{\OI}{6300}}
\newcommand\SIIlam{\linelam{\SII}{6731}}
\newcommand\SIIlamshort{\linelam{\SII}{6716}}
\newcommand\SIIlamboth{\linelam{\SII}{6716,6731}}
\newcommand\SIIIlam{\linelam{\SIII}{6312}}
\newcommand\NIIlam{\linelam{\NII}{6584}}
\newcommand\OIIIlam{\linelam{\OIII}{5007}}
\newcommand\Halam{\linelam{\Ha}{6563}}
\newcommand\Ha{\ensuremath{\mathrm{H}\alpha}}
\newcommand\Hb{\ensuremath{\mathrm{H}\beta}}
\newcommand\OI{[\ion{O}{1}]}
\newcommand\SII{[\ion{S}{2}]}
\newcommand\SIII{[\ion{S}{3}]}
\newcommand\OIII{[\ion{O}{3}]}
\newcommand\NII{[\ion{N}{2}]}
\newcounter{ionstage}
\renewcommand{\plainion}[2]{\setcounter{ionstage}{#2}%
  \ensuremath{\mathrm{#1\,\Roman{ionstage}}}}
\newcommand\plainOI{\mathrm{[\plainion{O}{1}]}}
\newcommand\plainOIII{\mathrm{[\plainion{O}{3}]}}
\newcommand\plainNII{\mathrm{[\plainion{N}{2}]}}
\newcommand\plainSIII{\mathrm{[\plainion{S}{3}]}}
\newcommand\plainSII{\mathrm{[\plainion{S}{2}]}}
\newcommand\HII{\ion{H}{2}}
\newcommand\HH[1]{HH~#1}
\newcommand\p{\ensuremath{^+}}
\newcommand\pp{\ensuremath{^{++}}}
\newcommand\EM{\ensuremath{E}}
\newcommand\emco{\ensuremath{\beta}}
\newcommand\A{\ensuremath{_{\mathrm{A}}}}
\newcommand\B{\ensuremath{_{\mathrm{B}}}}
\newcommand\avE{_{\scriptscriptstyle\!E}}
\newcommand\avS{_{\scriptscriptstyle\!S}}
\newcommand\GDH{\citetalias{2007AJ....133..952G}}%
\newcommand\DOH{\citetalias{2004AJ....127.3456D}}%

\newenvironment{wide}{}{}

\title{Velocity Structure in the Orion Nebula.\\ II. Emission Line
  Atlas of Partially Ionized\\ to Fully Ionized Gas}

\author{Ma.~T. Garc\'{\i}a-D\'{\i}az\altaffilmark{1}, W.~J.
  Henney\altaffilmark{2}, J.~A. L\'opez\altaffilmark{1}, and
  T. Doi\altaffilmark{3}}

\altaffiltext{1}{Instituto de Astronom\'{\i}a, Universidad
  Na\-cio\-nal Aut\'o\-no\-ma de M\'e\-xico, Ensenada}

\altaffiltext{2}{Centro de Radioastronom\'{\i}a y Astrof\'{\i}sica,
  Uni\-ver\-si\-dad Na\-cio\-nal Aut\'o\-no\-ma de M\'e\-xi\-co,
  Mo\-re\-lia}

\altaffiltext{3}{Japan Aerospace Exploration Agency, Tokyo, Japan}

\SetYear{2007}
\ReceivedDate{} 
\AcceptedDate{} 
\SetMSnumber{}

\resumen{%
  Presentamos un atlas\altaffilmark{4} de espectros tridimensionales
  (posici\'on-posici\'on-velocidad) de la Nebulosa de Ori\'on en
  l\'\i{}neas de emisi\'on \'opticas de una variedad de diferentes
  etapas de ionizaci\'on: \OIlam, \SIIlamboth, \NIIlam, \SIIIlam,
  \Halam, y \OIIIlam.  Estas transiciones nos dan informaci\'on punto
  a punto sobre la estructura f\'isica y cinem\'atica de la nebulosa a
  una resoluci\'on efectiva de $3\arcsec \times 2\arcsec \times
  10~\kms{}$, mostrando claramente el comportamiento a gran escala del
  gas ionizado y la presencia de fenomenos localizados tales como
  flujos colimados relacionados a objetos Herbig-Haro.  Como un
  ejemplo de la aplicaci\'on del atlas presentamos un an\'alisis
  estad\'istico de los anchos de las l\'ineas de \Ha{}, \OIII{} y
  \NII{}, que permiten una determinaci\'on de la temperatura
  electr\'onica media en la nebulosa de $(9200\pm
  400)$~K\@. Tambi\'en, a diferencia a trabajos anteriores,
  encontramos que no hay diferencia entre el ensanchamiento no
  t\'ermico de las l\'\i{}neas de recombinaci\'on y el de las
  l\'\i{}neas colisionales.}
 
\abstract{We present an atlas\altaffilmark{4} of three-dimensional
  (position-position-velocity) spectra of the Orion Nebula in optical
  emission lines from a variety of different ionization stages:
  \OIlam, \SIIlamboth, \NIIlam, \SIIIlam, \Halam, and \OIIIlam. These
  transitions provide point to point information about the physical
  structure and kinematics of the nebula at an effective resolution of
  $3\arcsec \times 2\arcsec \times 10~\kms{}$, clearly showing the
  large scale behavior of the ionized gas and the presence of
  localized phenomena such as Herbig-Haro outflows.  As an example
  application of the atlas, we present a statistical analysis of the
  widths of the \Ha{}, \OIII{}, and \NII{} lines that permits a
  determination of the mean electron temperature in the nebula of
  $(9200\pm 400)$~K\@. We also find, in contradiction to previous
  claims, that the non-thermal line broadening is not significantly
  different between recombination lines and collisional lines.  }

\altaffiltext{4}{Publicly available from
  \protect\url{http://www.astrosmo.unam.mx/~w.henney/orionatlas}.}

\addkeyword{H II regions}
\addkeyword{ISM: Herbig-Haro objects}
\addkeyword{ISM: Herbig-Haro objects}
\addkeyword{ISM: individual: Orion nebula}
\addkeyword{techniques: spectroscopy}
\addkeyword{turbulence}

\shortauthor{Garcia-Diaz et al.}

\shorttitle{Emission Lines Atlas of the Orion Nebula}

\fulladdresses{
\item T. Doi: Japan Aerospace Exploration Agency, 7-44-1
  Jindaiji-Higashi-machi, Chofu-shi, Tokyo 182-8522, Japan
  (takao.doi1@jsc.nasa.gov)
\item Ma.\ T. Garc\'{\i}a-D\'{\i}az and J. A. L\'opez: Instituto de
  Astronom\'{\i}a, Universidad Nacional Aut\'onoma de M\'exico, Campus
  Ensenada, Apartado Postal 22860, Ensenada, Baja California, M\'exico
  (tere,jal@astrosen.unam.mx)
 \item W. J. Henney: Centro de Radioastronom\'{\i}a y
   Astrof\'{\i}sica, Universidad Nacional Aut\'onoma de M\'exico,
   Campus Morelia, Apartado Postal 3--72, 58090 Morelia, Michoac\'an,
   M\'exico (w.henney@astrosmo.unam.mx)
 }

\listofauthors{Ma.\ T. Garc\'{\i}a-D\'{\i}az, W. J. Henney, J. A.
   L\'opez, \& T. Doi}

\indexauthor{Garc\'{\i}a-D\'{\i}az, Ma.\ T.}  
\indexauthor{Henney, W.~J.}  
\indexauthor{L\'opez, J.~A.}  
\indexauthor{Doi, T.}

\begin{document}
\maketitle

\section{Introduction}
\label{sec:introduction}

The Orion nebula, M42, is the nearest and best studied high-mass
star-forming region in our galaxy \citep[see][and references
therein]{2001ARA&A..39...99O}. It is being ionized by a compact group
of high-mass stars known as the Trapezium, which was created in the
first of three recent epsiodes of star formation that have occured
within a region of size one-tenth of a parsec at the center of the
Orion Nebula Cluster. The other two star formation episodes,
associated with the BN/KL and Orion South regions, are still ongoing
and are ocurring in dense molecular gas that lies behind the ionized
nebula. The O7~V star \thC{} \citep{2006A&A...448..351S} is the most
luminous and hottest star of the Trapezium and it is responsible for
the main ionization front. Many morphological features in Orion that
are apparent in emission lines, such as arcs and filaments are
associated with collimated gas flows from star-forming activity, such
as Herbig-Haro (HH) objects \citep[e.g.,][]{2001ARA&A..39..403R}.
Jets, bright bars, proplyds and a myriad of other features are
conspicuous in optical emission lines making of Orion an ideal
laboratory for studying photoionization and hydrodynamic
phenomena. Previous high spectral resolution studies over restricted
regions in Orion \citep[e.g.,][]{1988ApJS...67...93C,
  1993ApJ...409..262W, 1998MNRAS.295..401E, 2000ApJS..129..229B}
indicate the presence of complex line profiles; the changing
complexity of the emission line profiles over distinct regions of
Orion can now be fully appreciated and clearly identified with various
phenomena over the entire region covered by our observations.
   
In this paper we present a comprehensive atlas of high-resolution
echelle slit spectra in multiple emission lines, covering a $3'\times
5'$ region in the center of the Orion nebula. The atlas combines
results from two separate datasets, one obtained at Kitt Peak National
Observatory (KPNO), and the other from the Observatorio Astron\'omico
Nacional at San Pedro M\'artir, B.C., Mexico (SPM). Between them,
these datasets, which have comparable spatial and spectral resolution,
span a wide range of ionization stages, going from \OIlam, which
traces partially ionized gas at the ionization front, up to \OIIIlam,
which traces the most highly ionized gas in the interior of the
nebula. Figures and data from the atlas are publicly available from
the web page \url{http://www.astrosmo.unam.mx/~w.henney/orionatlas}.

\defcitealias{2007AJ....133..952G}{Paper~I}%
\defcitealias{2004AJ....127.3456D}{DOH04}%
Various aspects of the data presented here have been discussed in
previous papers: \citet[hereafter Paper~I]{2004AJ....127.3456D}
concentrated on the high-velocity features associated with stellar
jets; \citet{2005ApJ...627..813H} analysed the large-scale nebular
structure and kinematics in terms of numerical simulations of
champagne flows; \citet[hereafter Paper~I]{2007AJ....133..952G}
investigated the global structures found in low-velocity emission from
low-ionization gas; \citet{2007AJ....133.2192H} analysed the structure
and dynamics of the low-excitation jet, \HH{528}. However the full set
of individual long-slit spectra are shown here for the first time and
the usefulness and applicability of this comprehensive data set goes
beyond the analysis of the previous papers. As example applications of
the atlas data, we calculate velocity moment maps of the nebula in
each emission line and present a comprehensive analysis of the line
profile statistics.

The structure of the paper is as follows. In \S~\ref{sec:atlas} we
present the atlas of slit spectra. In \S~\ref{sec:moment-maps} we
summarise the spatial variations in line profiles across the face of
the nebula by means of maps of integrated line quantities, such as
centroid velocity and linewidth. In \S~\ref{sec:velocity-statistics}
we analyse the statistical properties of the velocity field in the
nebula, including correlations between the kinematics in different
emission lines. In \S~\ref{sec:deriv-mean-electr} we derive the
nebular electron temperature from the difference in linewidths of
\Ha{} and \OIII{}. In \S~\ref{sec:discussion} we discuss our results
in the light of previous work. in \S~\ref{sec:conclusions} we present
our conclusions.


\section{The Atlas}
\label{sec:atlas}

\begin{sidewaysfigure*}[p]\centering
\includegraphics[angle=270]{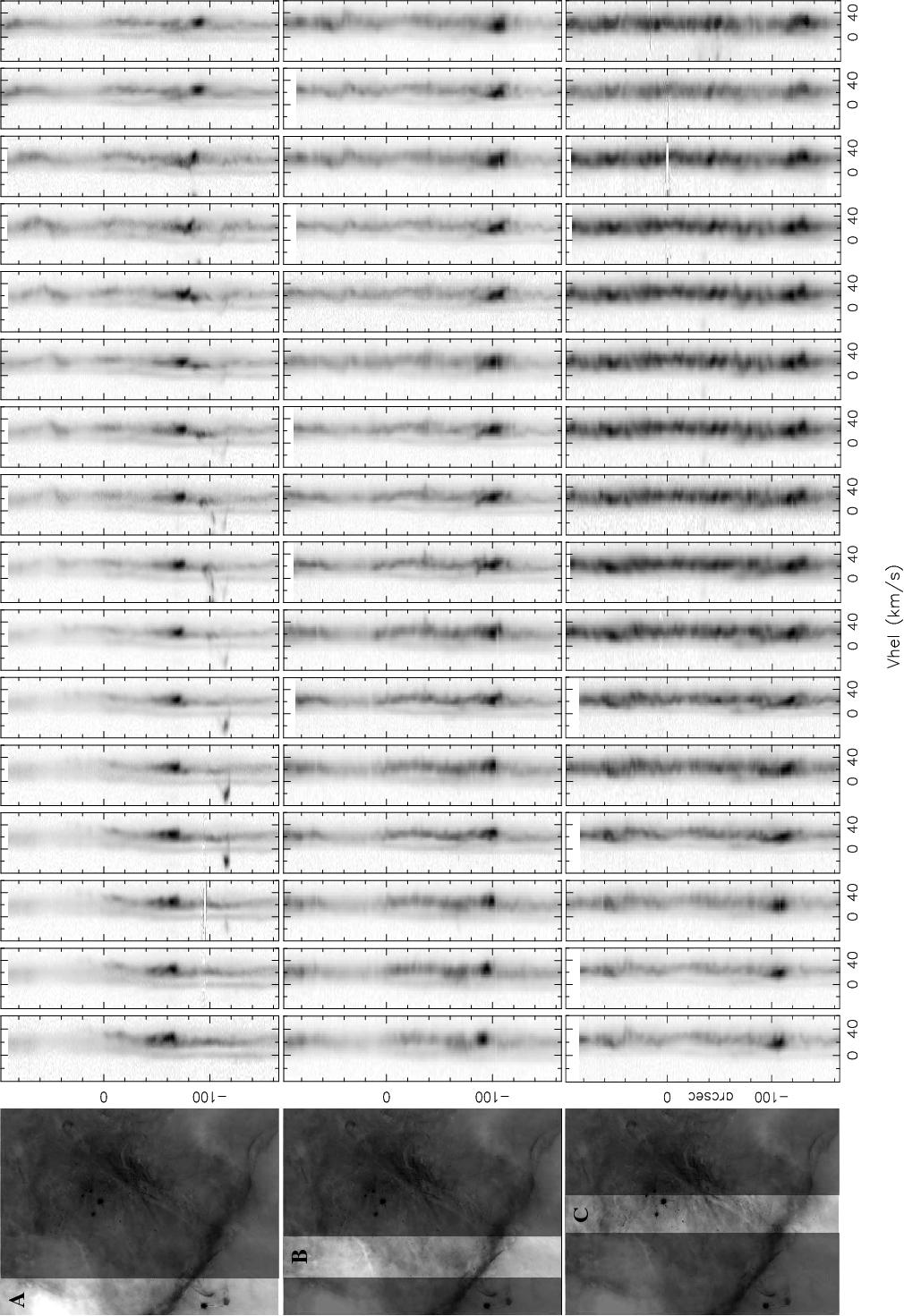}
\caption{\SIIlam{} individual long-slit spectra are presented as
  position-velocity (P-V) arrays. For every P-V array the vertical
  axis covers a range of 250\arcsec{} in declination, while the
  horizontal axis from $-$40 to $+$60 \kms{} in heliocentric
  velocity. The strip of sky corresponding to the slits in each row is
  highlighted on an HST image of Orion, shown in the first column. The
  spectra in the top row (A) are from Kitt Peak (resolution, R,
  $\approx$ 8 \kms), exept for P-V's 11, 15 \& 16 that are from SPM (R
  $\approx$ 6 \kms). All other spectra in the middle and bottom rows
  (B \& C) are from SPM (R $\approx$ 12 \kms)}
\label{fig:mosaico_a}
\end{sidewaysfigure*}

\begin{sidewaysfigure*}[p]\centering
  \includegraphics[angle=270]{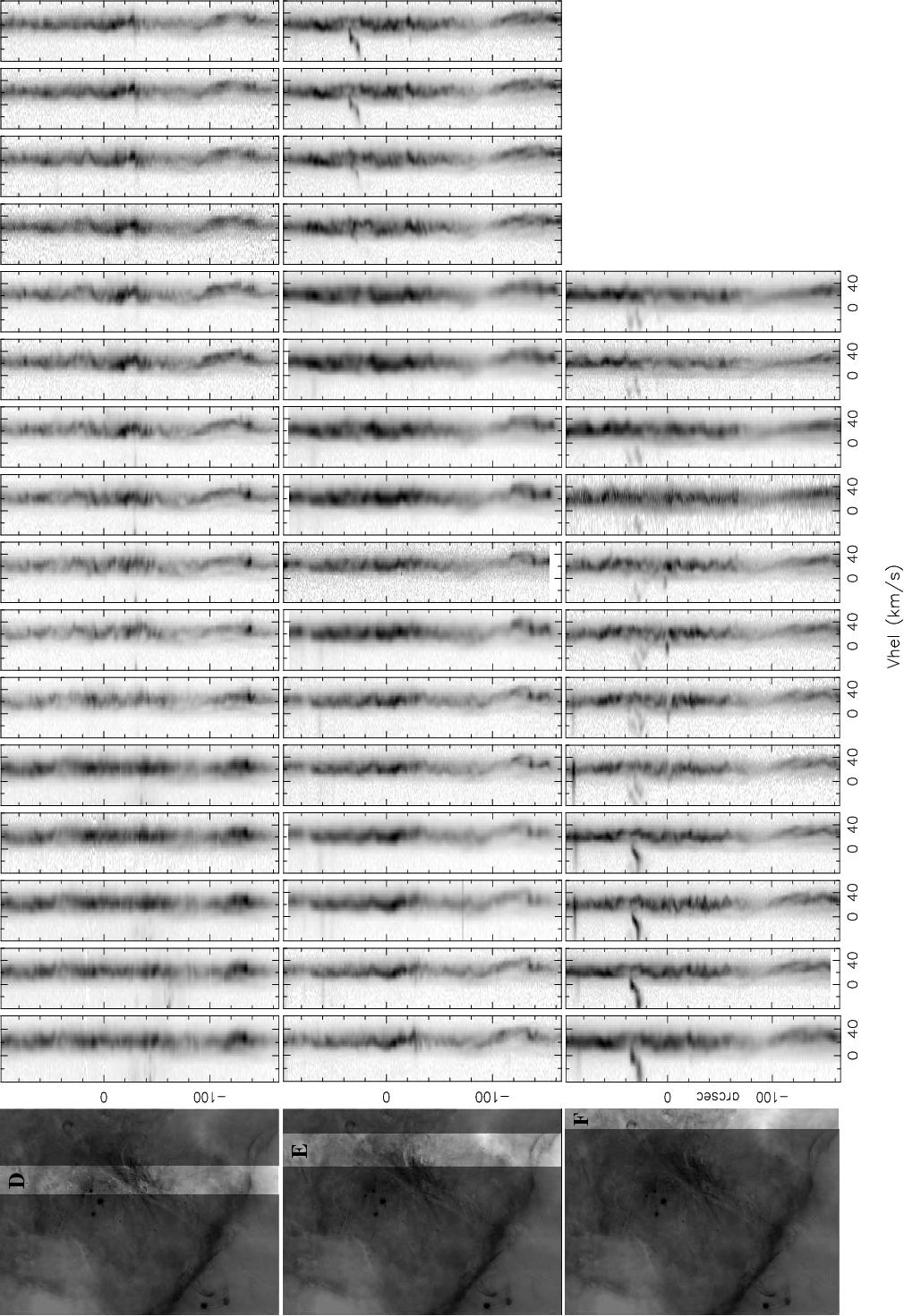}
  \caption{As Figure 1. In strip D the P-V arrays 6 to 16 are
    from KPNO, the rest are from SPM. In strip E, P-V arrays 1,2,6,13
    to 16 are from KPNO, the rest are from SPM. In strip F, P-V arrays
    3 and 5 to 9 are from KPNO, the rest are from SPM }
  \label{fig:mosaico_b} 
\end{sidewaysfigure*}

\begin{sidewaysfigure*}[p]\centering
  \includegraphics[angle=270]{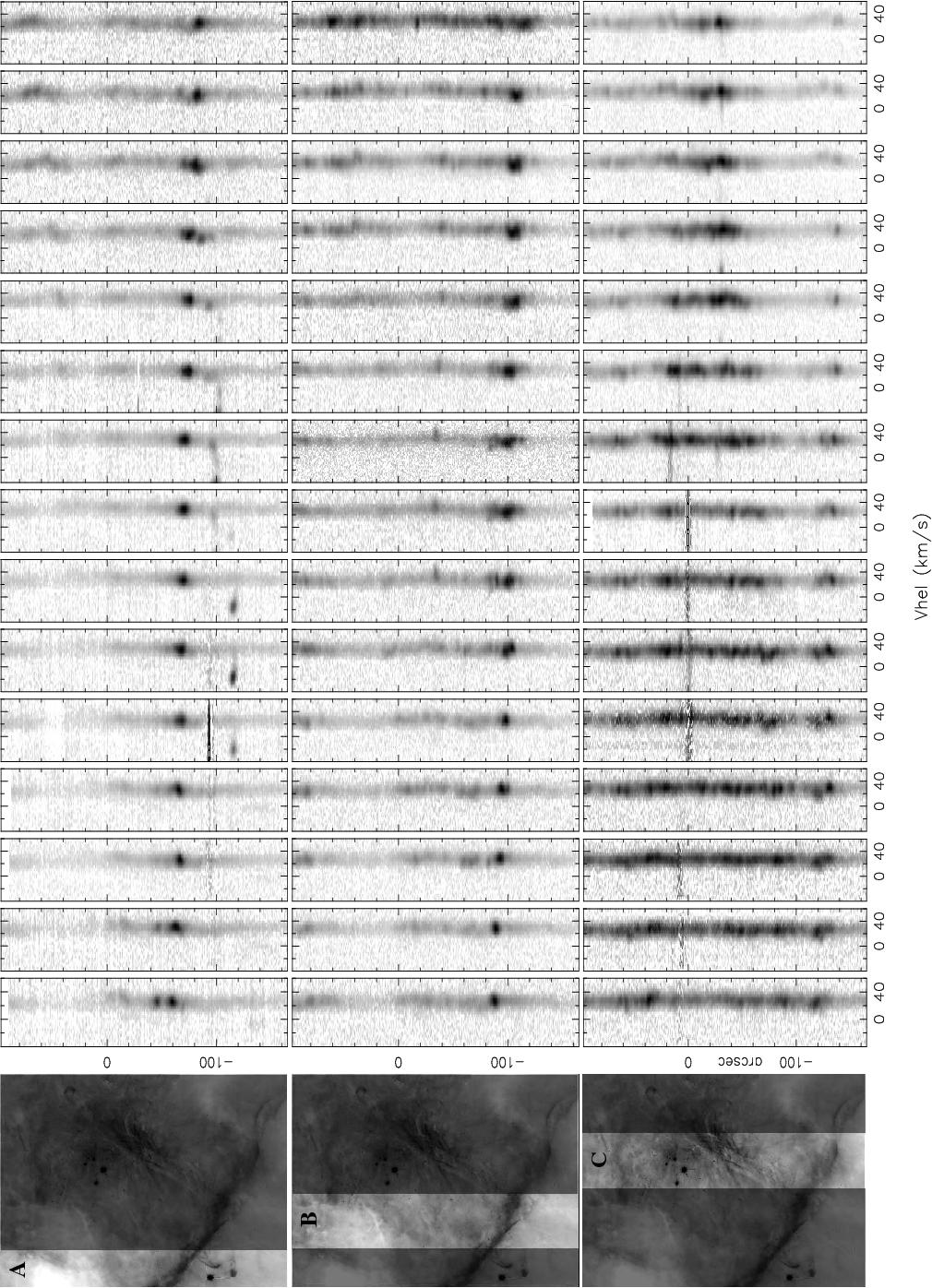}
  \caption{As in Figure 1 but for \OIlam{}. All the
    P-V arrays here are from SPM obtained at R $\approx$ 12 \kms,
    except P-V 9 in strip B, obtained at R $\approx$ 6 \kms }
  \label{fig:mosaico_c} 
\end{sidewaysfigure*}

\begin{sidewaysfigure*}[p]\centering
  \includegraphics[angle=270]{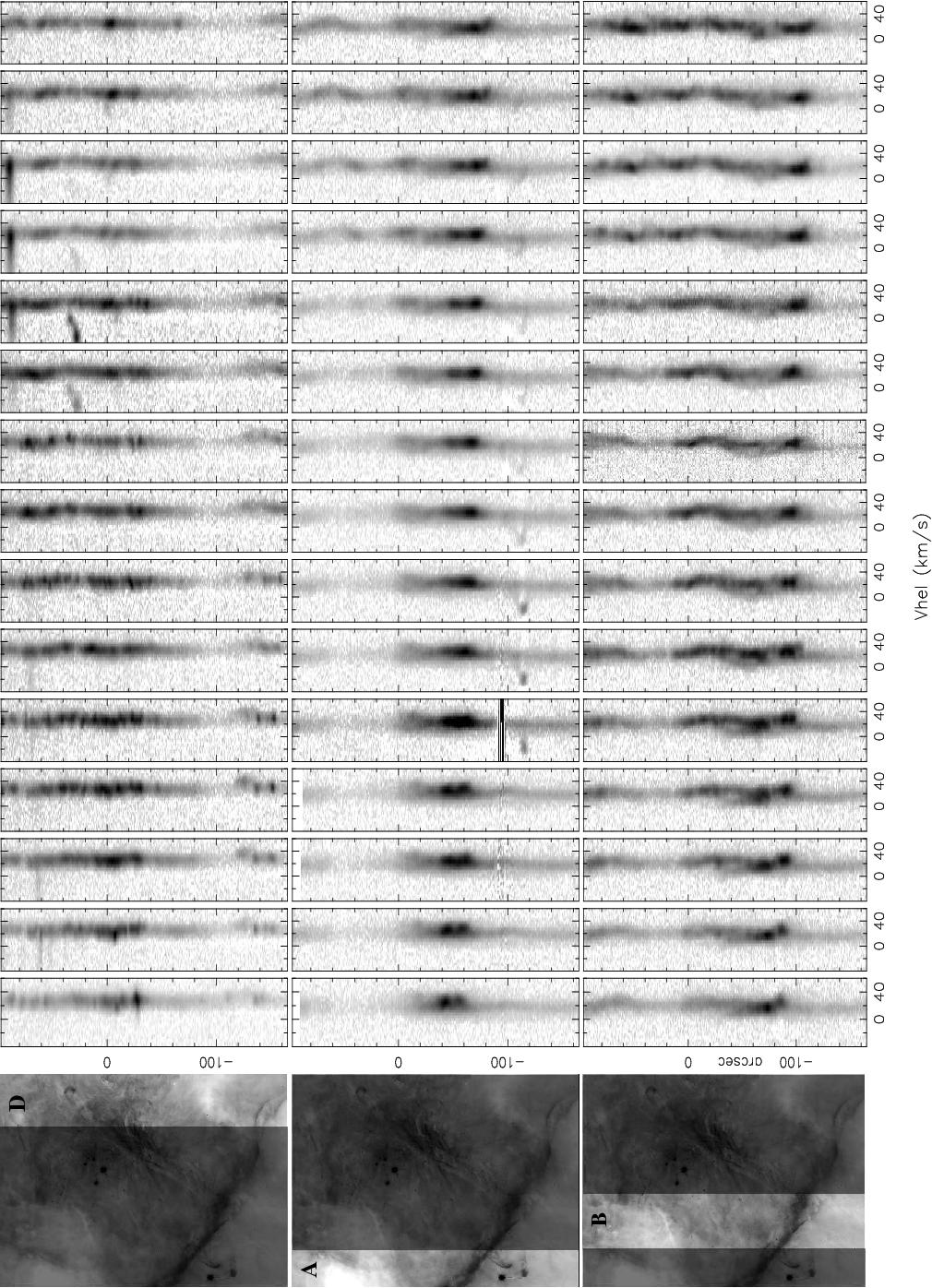}
  \caption{Strip D, top row, shows continuation of \OIlam{} mapping,
    as in Figure 3. The P-V arrays in the middle and bottom rows (A \&
    B) correspond to emission from \SIIIlam{}, all the
    spectra are from SPM obtained at R $\approx$ 12 \kms, except P-V 9
    in the bottom row, obtained at R $\approx$ 6 \kms }
  \label{fig:mosaico_d} 
\end{sidewaysfigure*}

\begin{sidewaysfigure*}[p]\centering
  \includegraphics[angle=270]{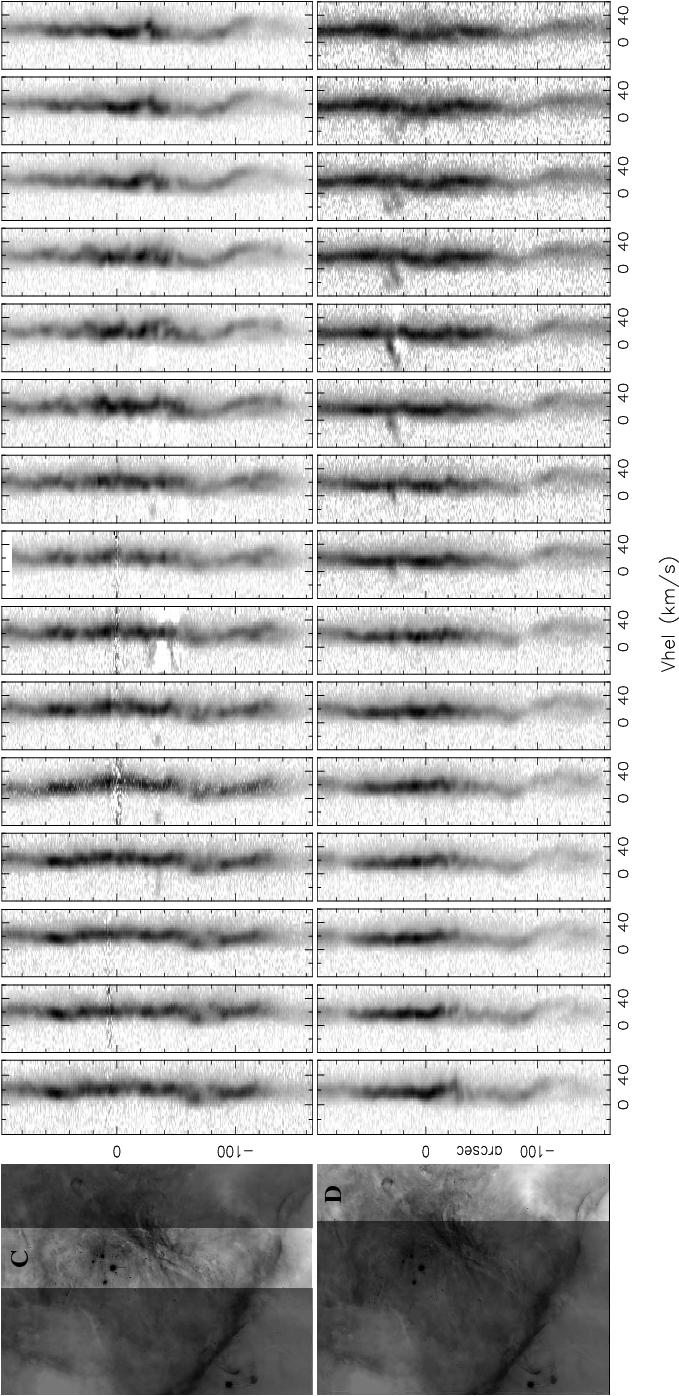}  
  \caption{P-V arrays of \SIIIlam{}, continued from
    Figure 4}
  \label{fig:mosaico_e} 
\end{sidewaysfigure*}

\begin{sidewaysfigure*}\centering
  \includegraphics[angle=270]{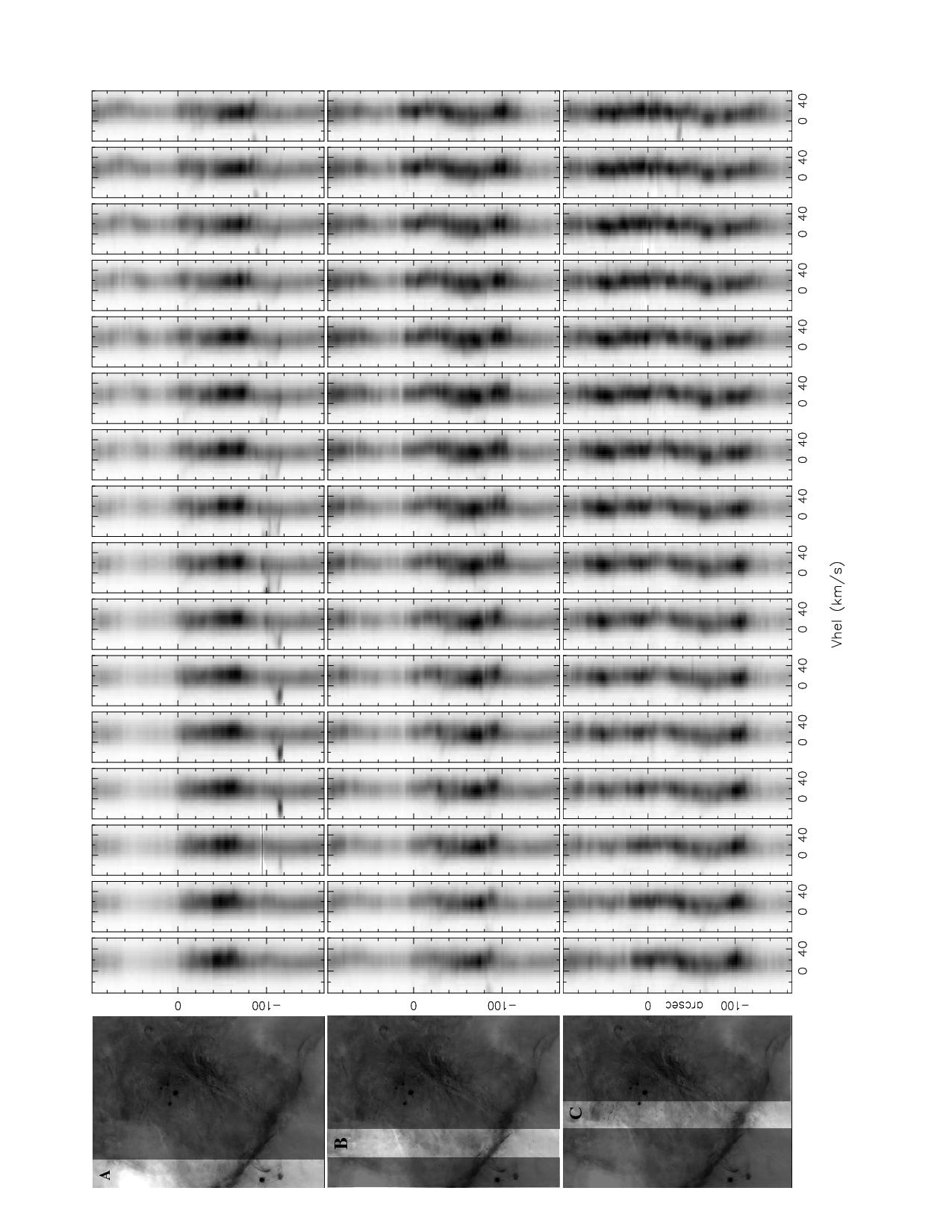}
  \caption{As Figure 1 but for \Halam{}.}
  \label{fig:mosaico_ha01} 
\end{sidewaysfigure*}

\begin{sidewaysfigure*}\centering
\includegraphics[angle=270]{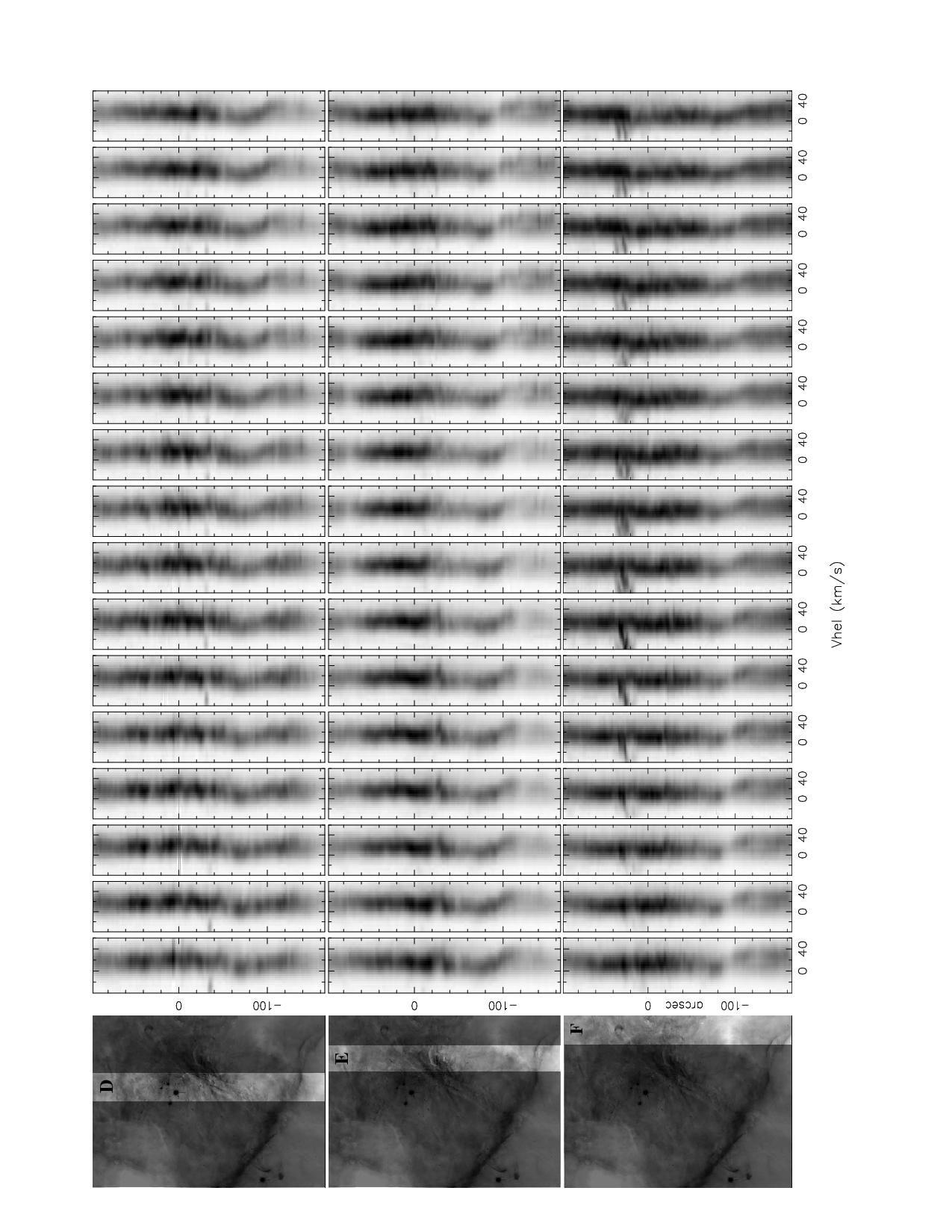}
  \caption{As Figure 1 but for \Halam{}, continued.}
  \label{fig:mosaico_ha02} 
\end{sidewaysfigure*}

\begin{sidewaysfigure*}\centering
  \includegraphics[angle=270]{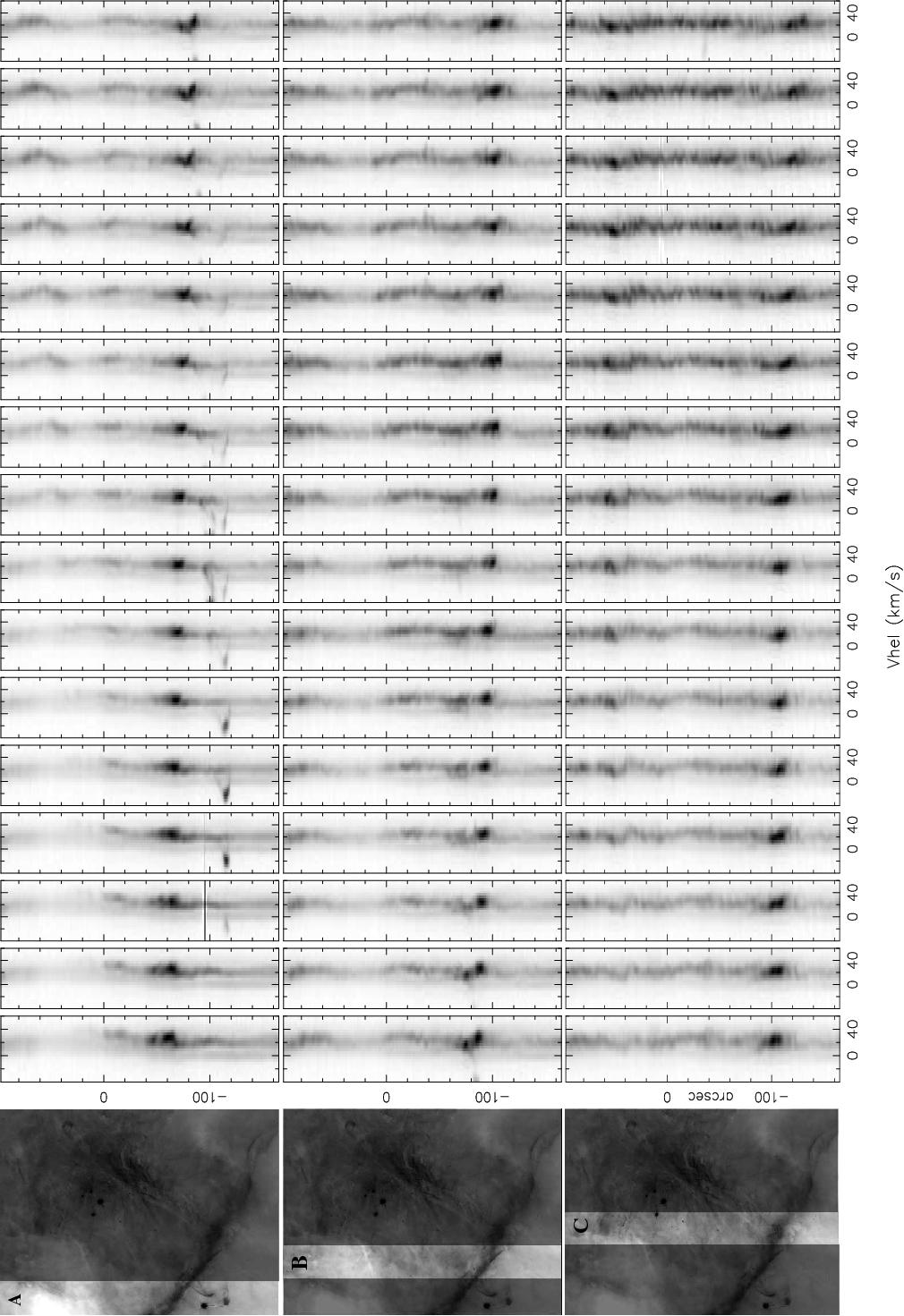}
  \caption{As Figure 1 but for \NIIlam{}.}
  \label{fig:mosaico_nii01} 
\end{sidewaysfigure*}

\begin{sidewaysfigure*}\centering
  \includegraphics[angle=270]{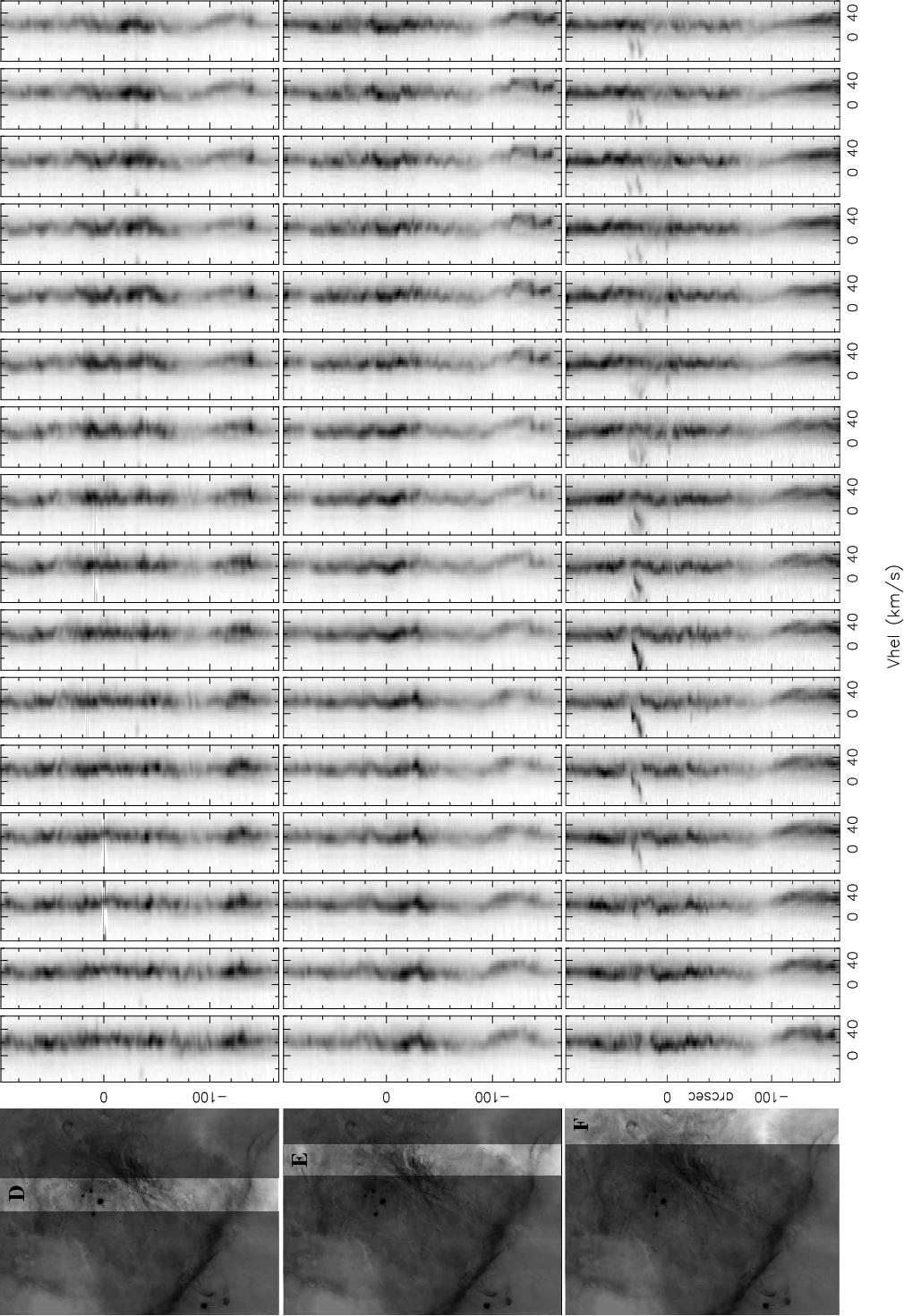}
  \caption{As Figure 1 but for \NIIlam{}, continued.}
  \label{fig:mosaico_nii02} 
\end{sidewaysfigure*}

\begin{sidewaysfigure*}\centering
  \includegraphics[angle=270]{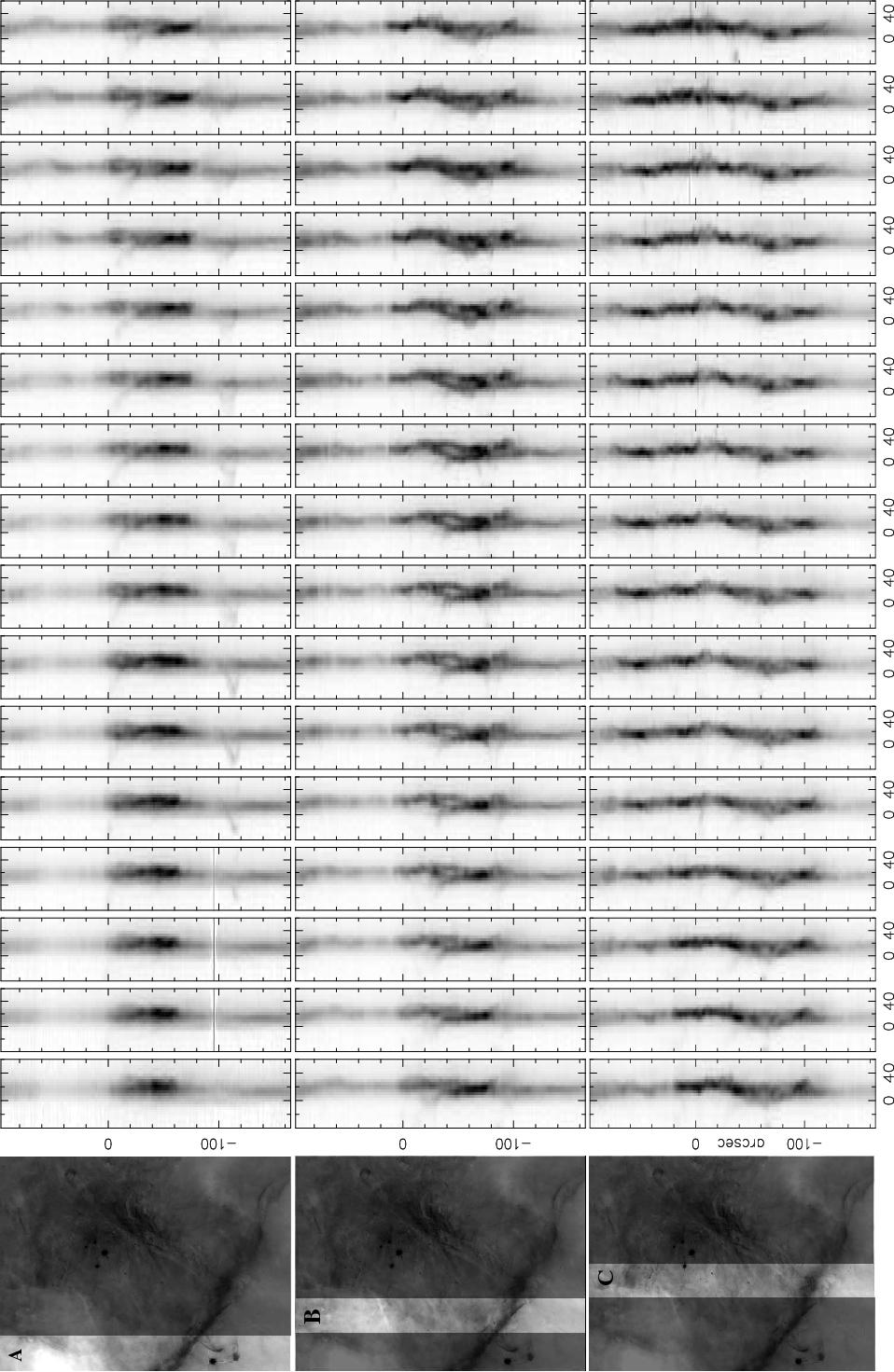}
  \caption{As Figure 1 but for \OIIIlam{}.}
  \label{fig:mosaico_oiii01} 
\end{sidewaysfigure*}

\begin{sidewaysfigure*}\centering
  \includegraphics[angle=270]{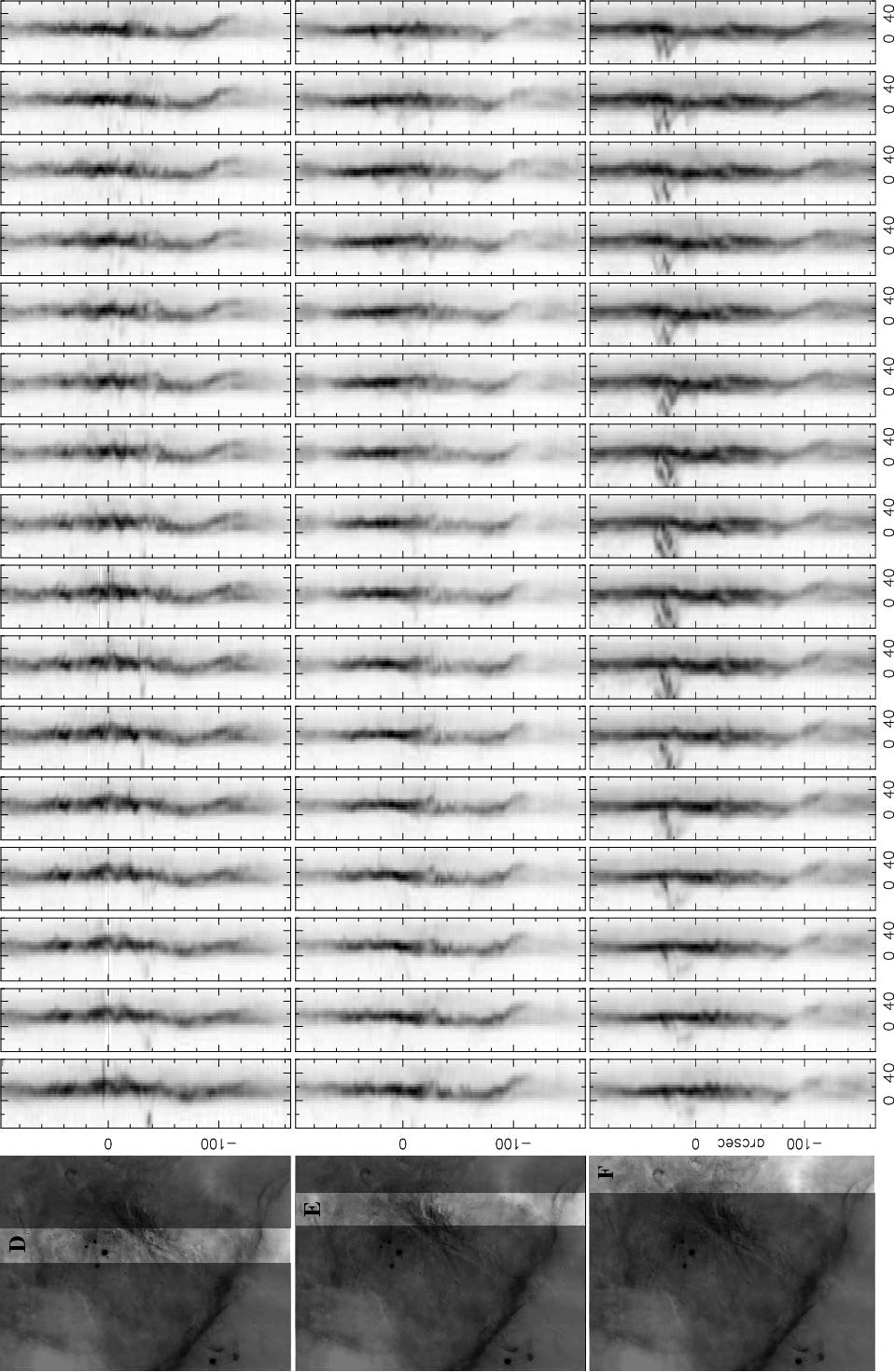}
  \caption{As Figure 1 but for \OIIIlam{}, continued.}
  \label{fig:mosaico_oiii02} 
\end{sidewaysfigure*}

\begin{table}
  \centering
  \caption{Emission lines in the atlas}
  \label{tab:atlas}
  \begin{tabular}{lccc}
    \toprule
    & \multicolumn{3}{c}{N-S slit positions} \\
    Line & KPNO & SPM & Total \\ \midrule
    \OIlam & \nodata & 60 & 60 \\
    \SIIlamboth & 37 & 55 & 92 \\
    \NIIlam & 96 & \nodata & 96 \\
    \SIIIlam & \nodata & 60 & 60 \\
    \Halam & 96 & \nodata & 96 \\
    \OIIIlam & 96 & \nodata & 96 \\
    \bottomrule
  \end{tabular}
\end{table}

The spectral line data used to construct the atlas come from
observations at two different observatories, KPNO and SPM\@. The
number of north-south oriented slits from each observatory that
contribute to the dataset for each emission line is given in
Table~\ref{tab:atlas}.

The SPM observations were obtained with the Manchester echelle
spectrometer \citep[MES;][]{2003RMxAA..39..185M} combined with the
f/7.8 focus on the 2.1 m San Pedro M\'artir telescope of the
Universidad Nacional Aut\'onoma de M\'exico, further details of the
observations are described in \GDH{}). The KPNO observations were
obtained with the echelle spectrometer attached to the f/8 Cassegrain
focus of the Mayall 4~m telescope at Kitt Peak National Observatory,
further details are given in \citet{2001ApJ...556..203O} and
\citet{2004AJ....127.3456D}.



\label{recalibration}
The SPM-MES spectra were obtained with a slit width of 150~$\mu$m
(2.0\arcsec) for a resolution of 12~\kms, although a few positions
were observed with a narrower, 70~$\mu$m (0.9\arcsec) slit and a
6~\kms, resolution. The slit length at SPM is 312\arcsec. For KPNO the
slit width is 130~$\mu$m (0.8\arcsec), the slit length 300\arcsec{}
and 8~\kms{} resolution. Further details of the data reduction steps
are given in \DOH{} and \GDH{}. Since the original analysis of \DOH{}
was oriented towards fast-moving jets and shocks, the tolerances
employed for their velocity calibration ($\simeq 4~\kms{}$) did not
need to be as tight as in \DOH{}. Furthermore, although \GDH{}
employed an E-W slit in order to tie together the velocity
calibrations of the multiple N-S slits, this was not done by
\DOH{}. Instead, it was assumed that the mean velocity, integrated
over each slit, was constant as a function of right ascension.

We have therefore internally re-calibrated each \DOH{} velocity cube
in two ways. First, we have applied small rectification corrections in
order to remove obvious discontinuities between adjacent slits, which
become apparent when examining the kinematics of slow-moving gas. For
\Ha{} and \NII{}, the affected slits are all those lying between $x =
-55$ and $x = 47$, whereas for \OIII{} no corrections were
necessary. A common velocity correction, which is linear in distance
along the slit, is applied to all the affected slits in each emission
line: $\delta V = a + b y$ ($\delta V$ is in \kms{}, while $x$ and $y$
are in arcsec with respect to \thC{}), with $a = -0.7$, $b = -0.013$
for \Ha{} and $a = -0.4$, $b = -0.015$ for \NII{}. The velocity
corrections to individual pixels vary between $\delta V \simeq
-3~\kms{}$ in the far south to $\delta V \simeq +1~\kms{}$ in the far
north. Although this procedure greatly reduced the discontinuities, it
did not remove them completely in the case of \Ha{}.

For the second recalibration, in order to correct large-scale trends
in the mean velocity as a function of right ascension, we have
obtained new observations in February 2007 through an E-W oriented
slit, positioned 22\arcsec{}S of \thC{}. These observations were
obtained with SPM-MES, using a 70~$\mu$m slit and an instrumental
configuration identical to that of \GDH{}, except for the filters
employed. The mean velocities of \NII{}, \Ha{}, and \OIII{} were
calculated as a function of RA along the slit, and these were compared
with the mean velocities extracted from the \DOH{} velocity cubes for
the same slit. For \NII{}, we found an rms difference $< 1~\kms{}$, so
no correction was applied. However, for \Ha{} and \OIII{}, systematic
discrepancies of $\sim 5~\kms{}$ are seen in the west of the
nebula. We have therefore fitted a third order polynomial to the
velocity differences as a function of RA and used it to correct each
slit of the \DOH{} cube.


In order to check the calibration of the absolute velocity scale for
each emission line, we have compared our results with those from
previous studies \citep{1980PASP...92...22B, 1988ApJS...67...93C,
  1993ApJ...409..262W,
  1999AJ....118.2350H, 2000ApJS..129..229B}. In all cases, we found
agreement to within \(2~\kms\). 


In summary, we estimate that over most of the \SII{}, \NII{}, \Ha{},
and \OIII{} maps, the relative velocities of each emission line are
accurate to $\sim 0.5~\kms{}$ over relatively short distances ($<
50''$) in the declination direction, but only accurate to $\simeq
1~\kms{}$ when comparing points lying in different slits or widely
separated in declination. For the \SIII{} and \OI{} lines, the low
signal-to-noise ratio over much of the map means that the accuracy is
much worse than this for individual spatial pixels, and therefore the
same velocity precision can only be retrieved by integrating the line
profile over an extended region and thus losing spatial
resolution. For the absolute value of the mean velocity of the
different lines, we also estimate an uncertainty of $1$--$2~\kms{}$.

Figures~\ref{fig:mosaico_a}--\ref{fig:mosaico_e} show the individual
long-slit spectra in the form of calibrated position-velocity (P-V)
arrays. The HST image of Orion at the left of the figures shows with a
highlighted strip the region covered by the sixteen slit positions
shown to the right.  The strip advances to the right (west) in the
following rows indicating the corresponding regions for the
spectra. Figures 1 \& 2 correspond to the \SIIlam{} emission line. We
do not show here the data for the \SIIlamshort{} component, although
the \SII{} line ratio has been used by \GDH{} to derive the spatially
resolved electron density for the whole region.  Strips A--F in
Figures~\ref{fig:mosaico_a} and~\ref{fig:mosaico_b} cover the full
survey area for \SII{}.  Figures~\ref{fig:mosaico_c} and
\ref{fig:mosaico_d} (Strips A--C) correspond to \OIlam{} and finally
Figures~\ref{fig:mosaico_d} (Strip D) and \ref{fig:mosaico_e} (Strips
A--B) correspond to \SIIlam{}.

The spectra are presented within a heliocentric velocity range of
$-40$ to $+60~\kms{}$ in all cases.  We choose this range because it
allows one to appreciate the complex structure of the nebula at these
relatively low velocities, although some high velocity features, as
those from HH jets fall outside the range displayed, but still can be
clearly identified. The declination scale and right ascension scale
set the zero referred with respect to \thC{}. The stellar continua
have been subtracted from the spectra. Note that the mean velocities
of the background molecular cloud and of the stars in the Orion Nebula
Cluster are both $\vhel \simeq 27 \pm 2~\kms$ and that the conversion
to ``local standard of rest'' (LSR) velocities is $V_\mathrm{LSR} =
\vhel - 18.1~\kms{}$.

\section{Maps of line profile parameters}
\label{sec:moment-maps}

\begin{figure*}
  \begin{wide}
    \centering
    \includegraphics{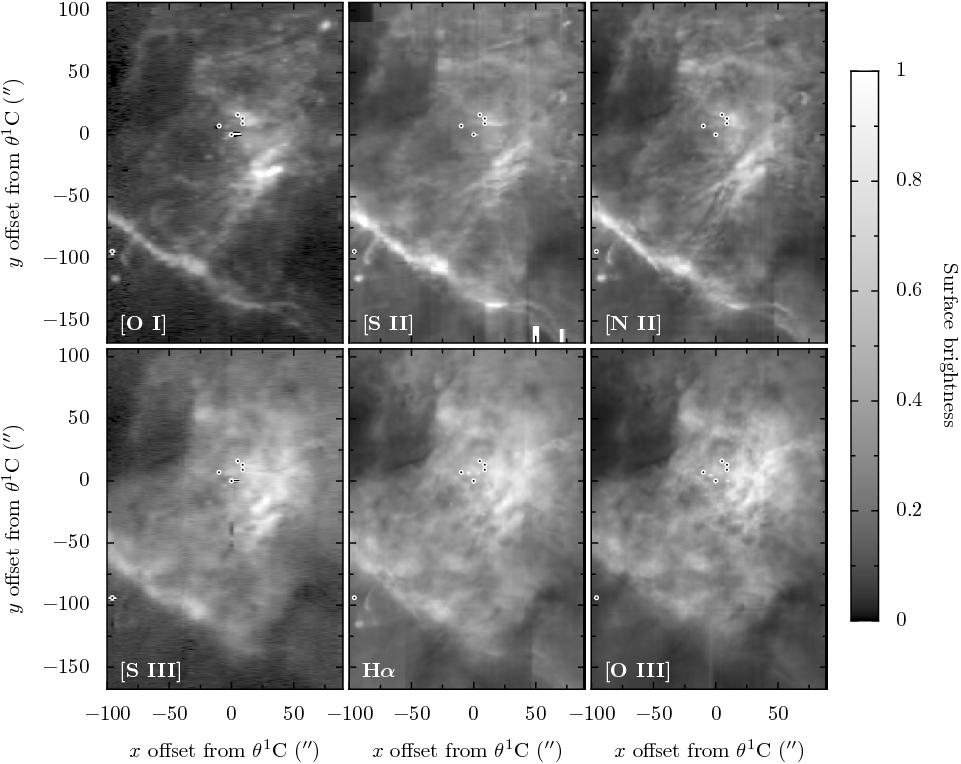}
  \end{wide}
  \caption{Reconstructed maps of emission line surface brightness,
    $S$, in the range $\vhel = -40$ to $\vhel = +70~\kms$.}
  \label{fig:moments-sum}
\end{figure*}
\begin{figure*}
  \begin{wide}
    \centering
    \includegraphics{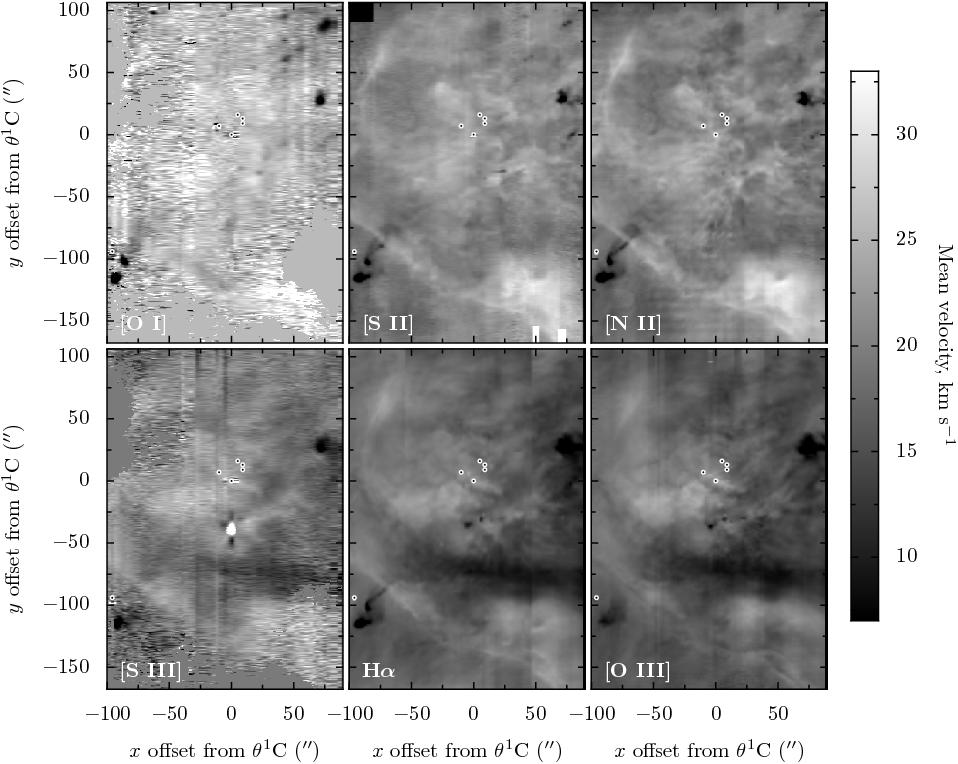}
  \end{wide}
  \caption{As Fig.~\ref{fig:moments-sum} but for the mean velocity,
    \vmean. }
  \label{fig:moments-mean}
\end{figure*}
\begin{figure*}
  \begin{wide}
    \centering
    \includegraphics{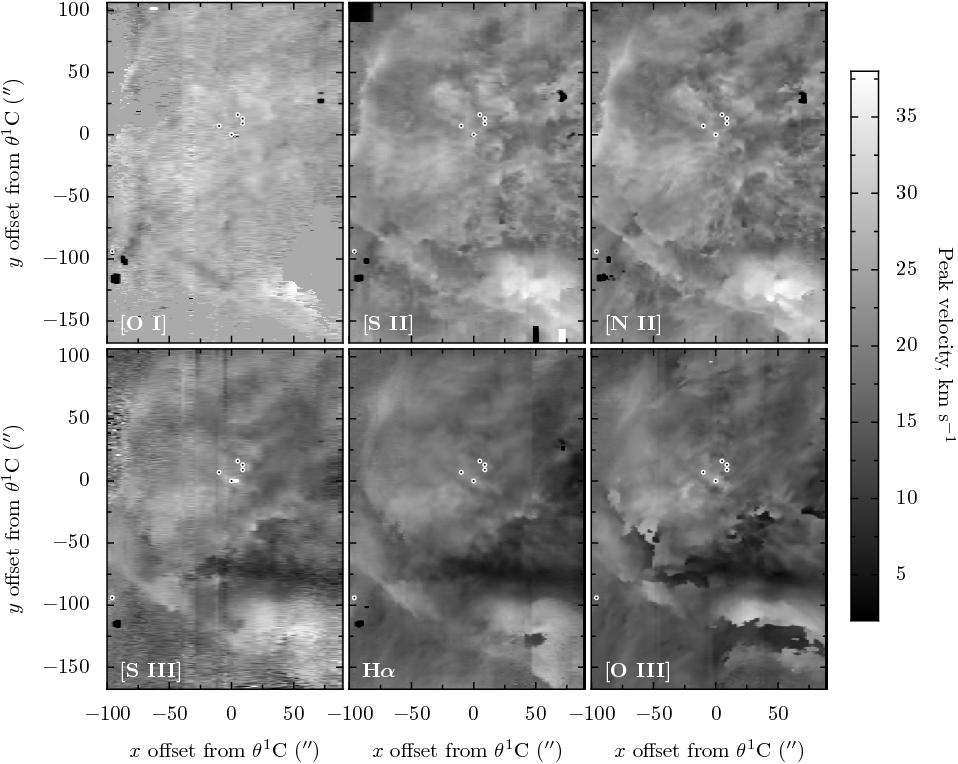}
  \end{wide}
  \caption{As Fig.~\ref{fig:moments-sum} but for the peak velocity,
    \vpeak.}
  \label{fig:moments-peak}
\end{figure*}
\begin{figure*}
  \begin{wide}
    \centering
    \includegraphics{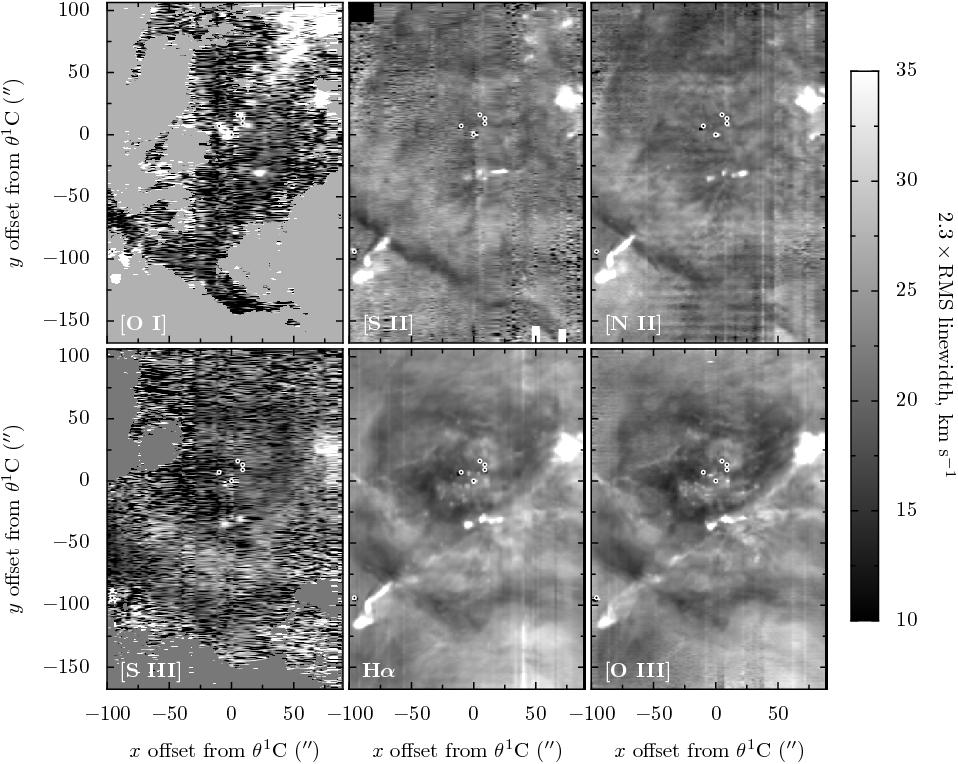}
  \end{wide}
  \caption{As Fig.~\ref{fig:moments-sum} but for the instrumentally
    and thermally corrected rescaled RMS line width, $ (8 \ln 2)^{1/2}
    \sigma$.}
  \label{fig:moments-sigma}
\end{figure*}
\begin{figure*}
  \begin{wide}
    \centering
    \includegraphics{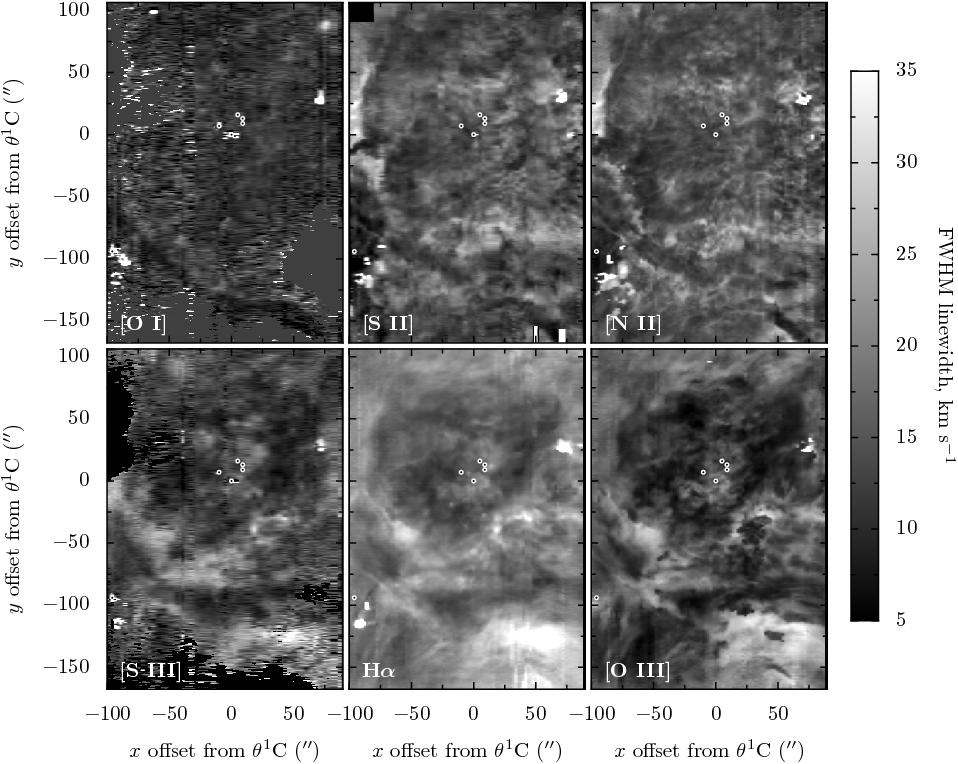}
  \end{wide}
  \caption{As Fig.~\ref{fig:moments-sum} but for the instrumentally
    and thermally corrected FWHM line width,
    $W$.}
  \label{fig:moments-fwhm}
\end{figure*}

In order to provide a global overview of the nebular structure and
kinematics, we show in Figures~\ref{fig:moments-sum} to
\ref{fig:moments-fwhm} maps of derived parameters of the line profile
of each emission line at each point in the nebula. These offer a
complementary approach to the isovelocity channel maps that have
already been published \citep{2004AJ....127.3456D,
  2007AJ....133..952G, 2007AJ....133.2192H}. The parameters shown are
total line surface brightness, $S$, mean velocity, \vmean,
root-mean-square (RMS) linewidth, $\sigma$, peak velocity, \vpeak, and
full width at half maximum (FWHM) linewidth, $W$. The first three of
these quantities can be defined in terms of the first three velocity
moments, $M_k$, of the line profiles $I(V)$:
\begin{equation}
  \label{eq:moments}
  M_k = \int_{V_1}^{V_2} V^k I(V) \, dV ,
\end{equation}
such that $S = M_0$, $\vmean = M_1/M_0$, and $\sigma^2 = (M_2/M_0) -
\vmean^2$. The limits of integration are chosen to be $V_1 = -40~\kms$
and $V_2 = +70~\kms$ for all lines, which represents a compromise
between being a wide enough range to include most emission of
interest, but not so wide as to introduce too much noise into the
maps. This is is a particular issue for $\sigma$, which is very
sensitive to noise in the far wings of the $I(V)$ profile. The
remaining quantities, \vpeak, and $W$ are directly measured from the
line profile. In order to determine \vpeak{} to a precision greater
than the size of the velocity pixels, we use parabolic interpolation
between the peak pixel and its two neighbours. Similarly, 
$W$ is calculated as the difference between the velocities of two
half-power points, which are found from linear interpolation between
adjacent pixels. In the case that the profile has more than two
half-power points, $W$ is defined as the difference between the
highest and lowest of these.

Figure~\ref{fig:moments-sum} shows the reconstructed surface
brightness maps in the six lines. Comparison with published direct
imaging \citep[e.g.,][]{1992ApJ...399..147P} shows generally good
agreement, although some artefacts remain from the process of
combining the individual spectra. These can easily be identified as
brightness discontinuities across vertical lines in the maps. The most
serious of these artefacts are in the \SII{} map at $x= +10$, $x=+45$,
in the \NII{} and \Ha{} maps at $x=+45$, in the \SIII{} map at $x=0$,
and in the \OIII{} map at $x=-5$. 

The mean and peak velocities are shown in
Figures~\ref{fig:moments-mean} and \ref{fig:moments-peak}, where it
can be seen that the two quantities are generally well-correlated for
a given line, although there are some important differences. The mean
velocity varies smoothly across the face of the nebula, whereas the
peak velocity shows frequent discontinuities, which are caused by one
particular line component becoming brighter than another. The mean
velocity is also more sensitive to high velocity components, such as
HH objects. This is particularly notable in the case of \OIII{}, where
the HH objects do not show up at all in the \vpeak{} image. Since
\SIII{} and \OI{} are very much weaker than the other lines, the
velocity maps are rather noisy, especially in the faint regions of the
nebula. We have therefore masked out from the velocity maps all
regions with a surface brightness below a certain threshold, which
appear as solid gray areas in the figures.

The non-thermal contributions to the RMS and FWHM line widths are
shown in Figure~\ref{fig:moments-sigma} and~\ref{fig:moments-fwhm}.
These are calculated by subtracting in quadrature the instrumental and
thermal widths from the raw measured widths, assuming a gas
temperature of $9100$~K (see \S~\ref{sec:deriv-mean-electr}). For a
Gaussian profile, the relationship between the FWHM and RMS widths is
$W = (8 \ln 2)^{1/2} \sigma \simeq 2.355 \sigma$. We therefore
multiply the values of $\sigma$ by 2.355 to allow direct comparison
between the $\sigma$ and $W$ maps. There is generally a much greater
difference between the $\sigma$ and $W$ maps of a given line than
between the maps of \vmean{} and \vpeak{}. The $\sigma$ maps are very
much dominated by high-velocity flows, which are much less apparent in
the $W$ maps. As with the \vpeak{} maps, the $W$ maps show
discontinuities between patches of the nebula with high and low line
widths. An exception is the \Ha{} line, where the increased thermal
broadening leads to a more Gaussian-like line profile and a closer
resemblance between the $W$ and $\sigma$ maps. As with the velocity
maps, we have masked out regions of low signal-to-noise ratio for the
\OI{} and \SIII{} lines.

An empirical description of the nebular structure and kinematics was
given in \S~3 of \GDH{} and will not be repeated here. The reader is
referred in particular to Fig.~7 of \GDH{} for orientation in
identifying nebular features mentioned in the remainder of the paper.

\section{One-point velocity statistics}
\label{sec:velocity-statistics}

\begin{table*}\centering
  \caption{Summary of velocity statistics}\label{tab:stats}
  \begin{tabular}{rrrrr}\toprule
    line  & \(\langle V \rangle\) & \(V_\mathrm{p}\) & \(\sigma^2\) & \(W\) \\
    \midrule
    \(\OIII\) & \(16.3 \pm 2.8\) & \(15.5 \pm 4.0\) & \(113 \pm 29\) & \(14.7 \pm 5.1 \) \\
    \(\Ha\) & \(16.8 \pm 3.0\) & \(16.2 \pm 4.1\) & \(195 \pm 28\) & \(18.6 \pm 5.1 \) \\
    \(\SIII\) & \(19.7 \pm 3.0\) & \(19.2 \pm 4.0\) & \(99 \pm 46 \) & \(14.6 \pm 4.4 \) \\
    \(\NII\) & \(20.5 \pm 2.9\) & \(21.1 \pm 3.7\) & \(106 \pm 24\) & \(15.4 \pm 3.7 \) \\
    \(\SII\) & \(21.2 \pm 2.4\) & \(21.8 \pm 3.2\) & \(114 \pm 21\) & \(15.2 \pm 3.5 \) \\
    \(\OI\) & \(25.7 \pm 3.4\) & \(26.3 \pm 2.5\) & \(106 \pm 65\) & \(11.5 \pm 2.6 \) \\
    \midrule
    \(\Ha-\OIII \) & \(0.5 \pm 1.4 \) & \(0.7 \pm 3.2 \) & \(82.2 \pm 11.8\) & \(3.9 \pm 3.0\) \\
    \(r       \) & \(0.889\) & \(0.700\) & \(0.917 \) & \(0.825\) \\
    \addlinespace
    \(\NII-\OIII\) & \(4.2 \pm 2.8 \) & \(5.6 \pm 4.5 \) & \(-7.0 \pm 36.8\) & \(0.7 \pm 6.5\) \\
    \(r       \) & \(0.511     \) & \(0.274     \) & \(0.287      \) & \(-0.067     \) \\
    \addlinespace
    \(\SII-\SIII\) & \(1.6 \pm 2.5 \) & \(2.6 \pm 3.3 \) & \(18.0 \pm 46.8\) & \(0.2 \pm 5.6\) \\
    \(r       \) & \(0.617     \) & \(0.588     \) & \(0.229      \) & \(0.065 \) \\
    \addlinespace
    \(\OI-\SII\) & \(4.4 \pm 3.0 \) & \(4.5 \pm 2.7 \) & \(-7.6 \pm 64.8\) & \(-3.7 \pm 4.2 \) \\
    \(r       \) & \(0.507     \) & \(0.580     \) & \(0.175      \) & \(0.078      \) \\
    \bottomrule
  \end{tabular}
\end{table*}

\newlength\panelwidth

\begin{figure*}
  \begin{wide}
    \centering
    \setlength\panelwidth{0.48\linewidth}
    \parbox[t]{\panelwidth}{(\textit{a})}\hfill
    \parbox[t]{\panelwidth}{(\textit{b})}\\
    \parbox[t]{\panelwidth}{\includegraphics{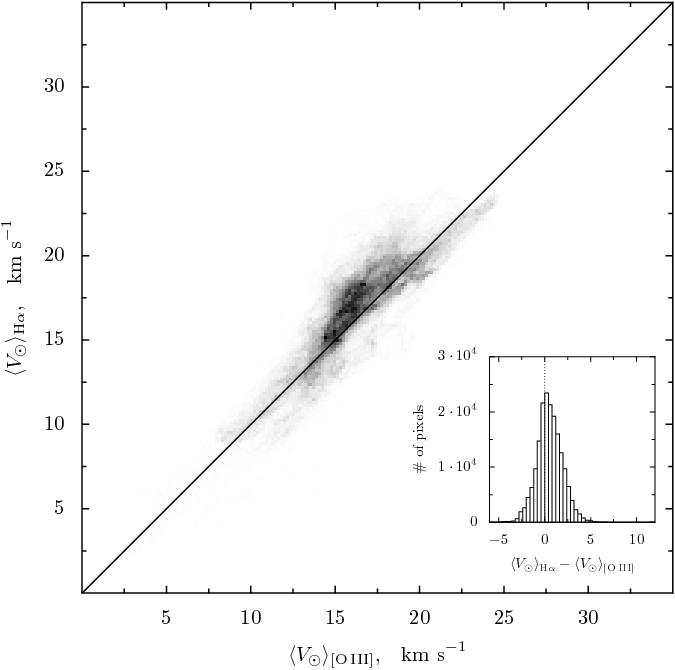}}\hfill
    \parbox[t]{\panelwidth}{\includegraphics{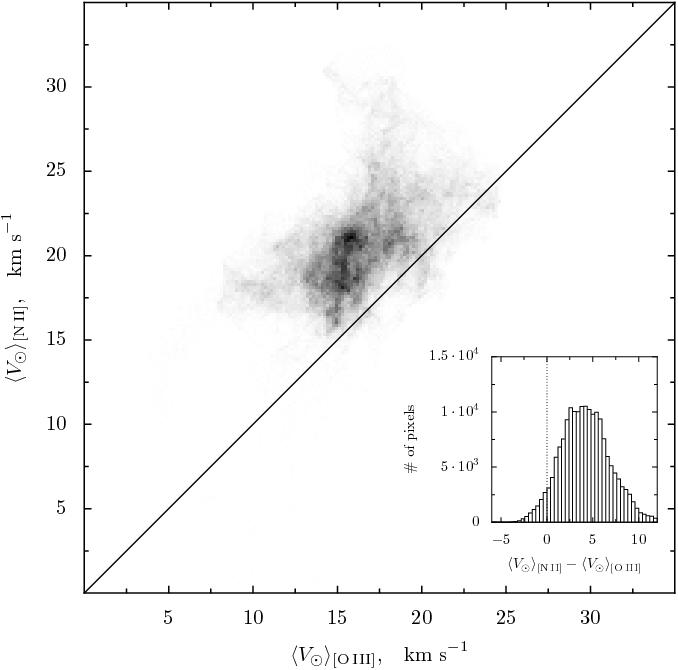}}
    \parbox[t]{\panelwidth}{(\textit{c})}\hfill
    \parbox[t]{\panelwidth}{(\textit{d})}\\
    \parbox[t]{\panelwidth}{\includegraphics{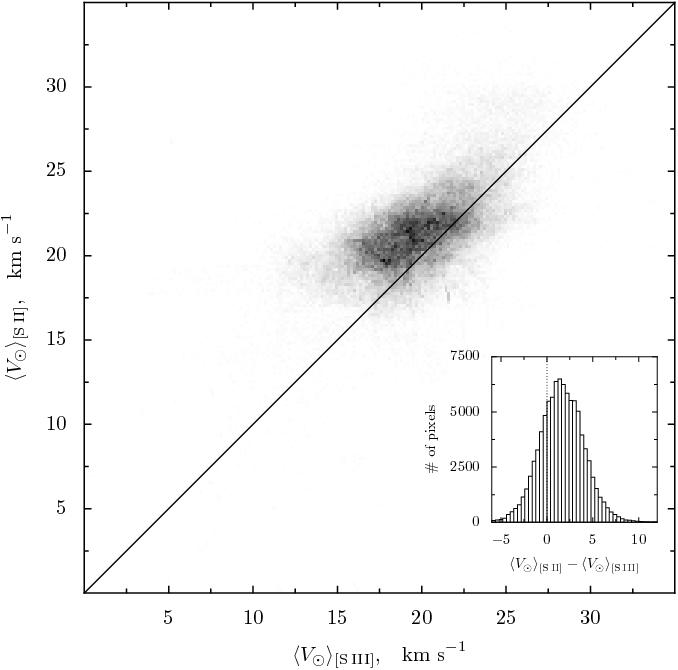}}\hfill
    \parbox[t]{\panelwidth}{\includegraphics{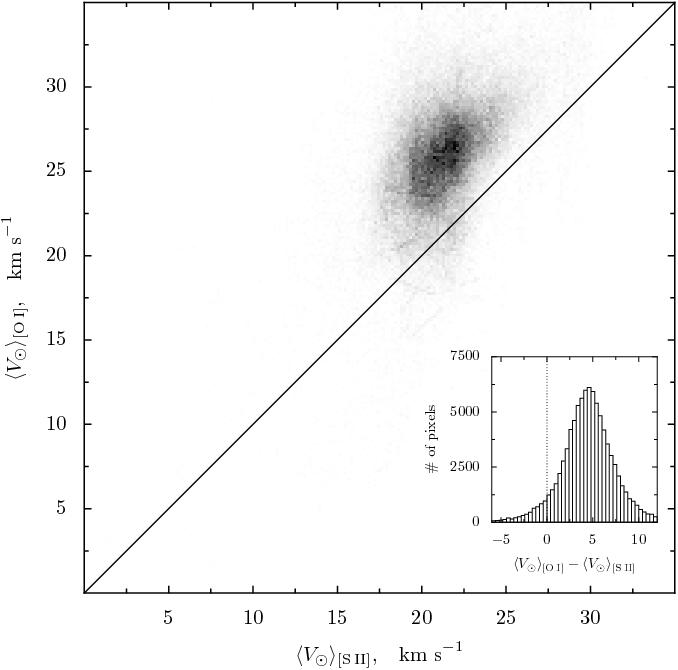}}
  \end{wide}
  \caption{Joint distribution of mean velocities, $\vmean$, of
    different emission lines, with inset box showing a histogram of
    differences in \vmean{}. In each case, the straight line shows the
    case of equal velocities in the two lines. (\textit{a})
    $\vmean_{\Ha}$ versus $\vmean_{\plainOIII}$. (\textit{b})
    $\vmean_{\plainNII}$ versus $\vmean_{\plainOIII}$ (\textit{c})
    $\vmean_{\plainSII}$ versus $\vmean_{\plainSIII}$. (\textit{d})
    $\vmean_{\plainOI}$ versus $\vmean_{\plainSII}$}
  \label{fig:meanvsmean}
\end{figure*}

\begin{figure*}
  \begin{wide}
    \centering
    \setlength\panelwidth{0.48\linewidth}
    \parbox[t]{\panelwidth}{(\textit{a})}\hfill
    \parbox[t]{\panelwidth}{(\textit{b})}\\
    \parbox[t]{\panelwidth}{\includegraphics{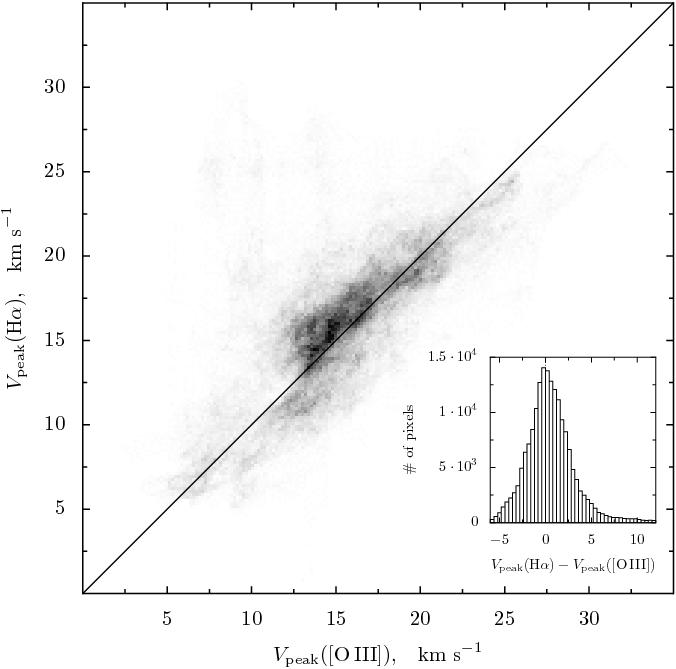}}\hfill
    \parbox[t]{\panelwidth}{\includegraphics{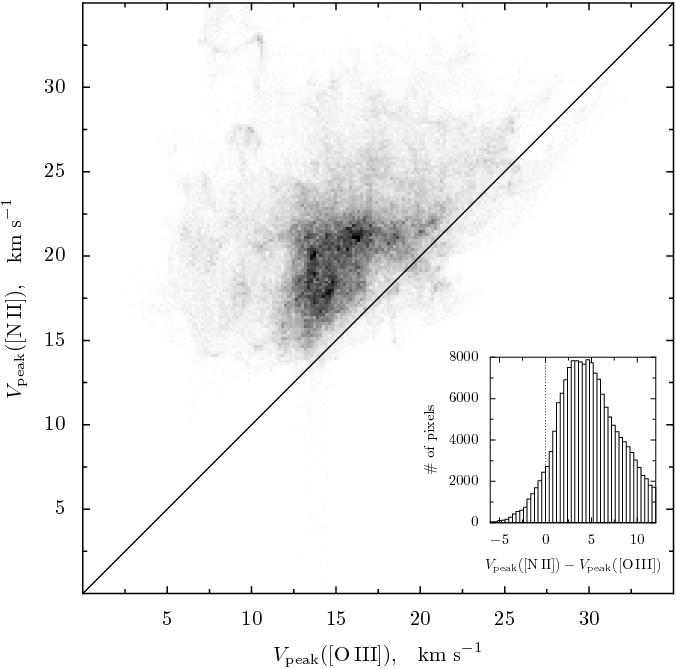}}
    \parbox[t]{\panelwidth}{(\textit{c})}\hfill
    \parbox[t]{\panelwidth}{(\textit{d})}\\
    \parbox[t]{\panelwidth}{\includegraphics{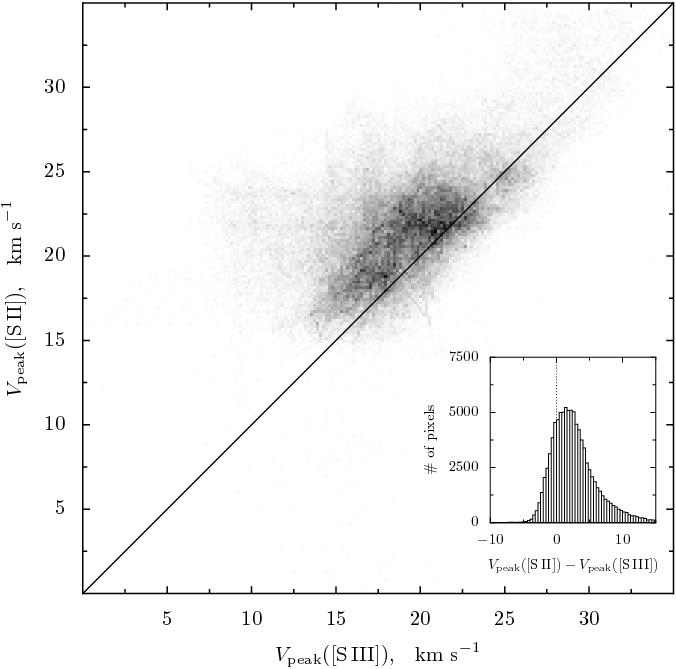}}\hfill
    \parbox[t]{\panelwidth}{\includegraphics{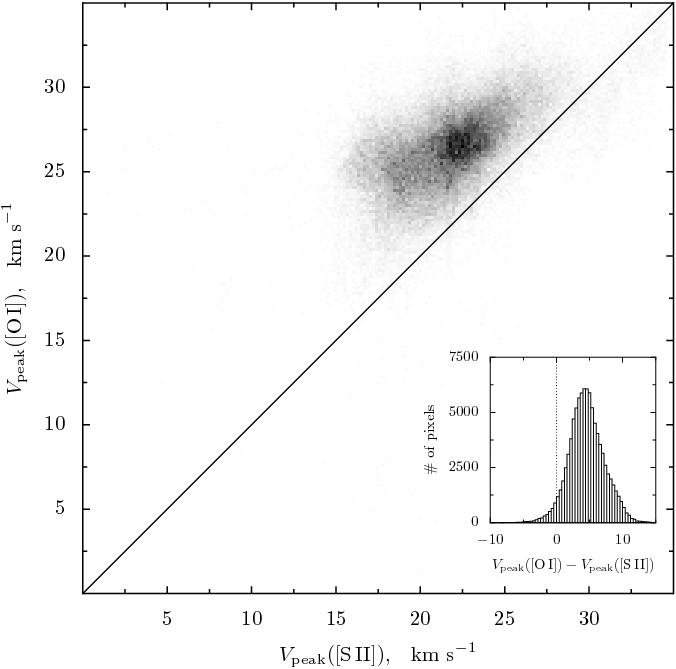}}
  \end{wide}
  \caption{Joint distribution of peak velocities, $\vpeak$, of
    different emission lines, with inset box showing a histogram of
    differences in \vpeak{}. In each case, the straight line shows the
    case of equal velocities in the two lines. (\textit{a})
    $\vpeak({\Ha})$ versus $\vpeak({\plainOIII})$. (\textit{b})
    $\vpeak({\plainNII})$ versus $\vpeak({\plainOIII})$ (\textit{c})
    $\vpeak({\plainSII})$ versus $\vpeak({\plainSIII})$. (\textit{d})
    $\vpeak({\plainOI})$ versus $\vpeak({\plainSII})$}
  \label{fig:peakvspeak}
\end{figure*}

\begin{figure*}
  \begin{wide}
    \centering
    \setlength\panelwidth{0.48\linewidth}
    \parbox[t]{\panelwidth}{(\textit{a})}\hfill
    \parbox[t]{\panelwidth}{(\textit{b})}\\
    \parbox[t]{\panelwidth}{\includegraphics{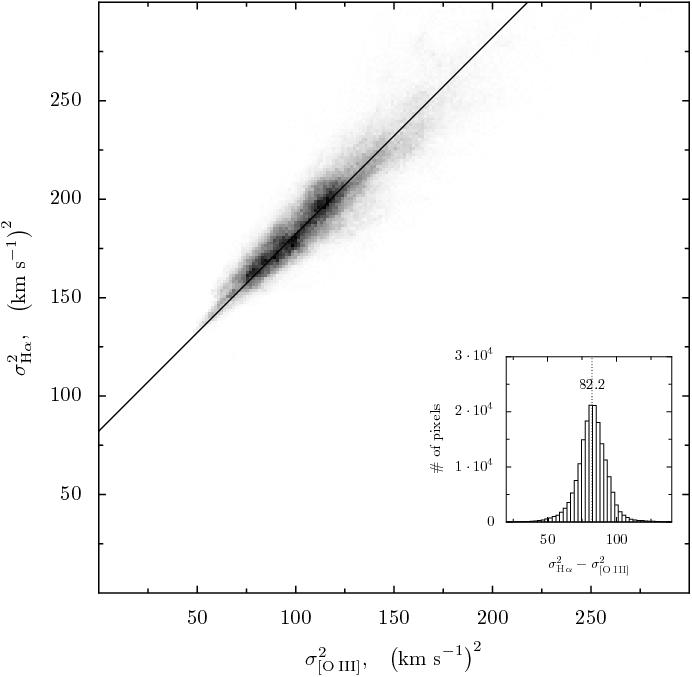}}\hfill
    \parbox[t]{\panelwidth}{\includegraphics{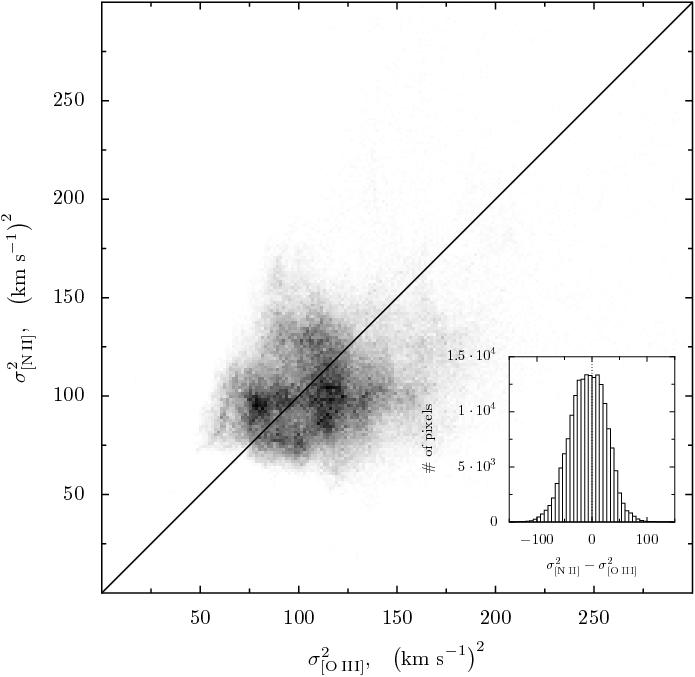}}
    \parbox[t]{\panelwidth}{(\textit{c})}\hfill
    \parbox[t]{\panelwidth}{(\textit{d})}\\
    \parbox[t]{\panelwidth}{\includegraphics{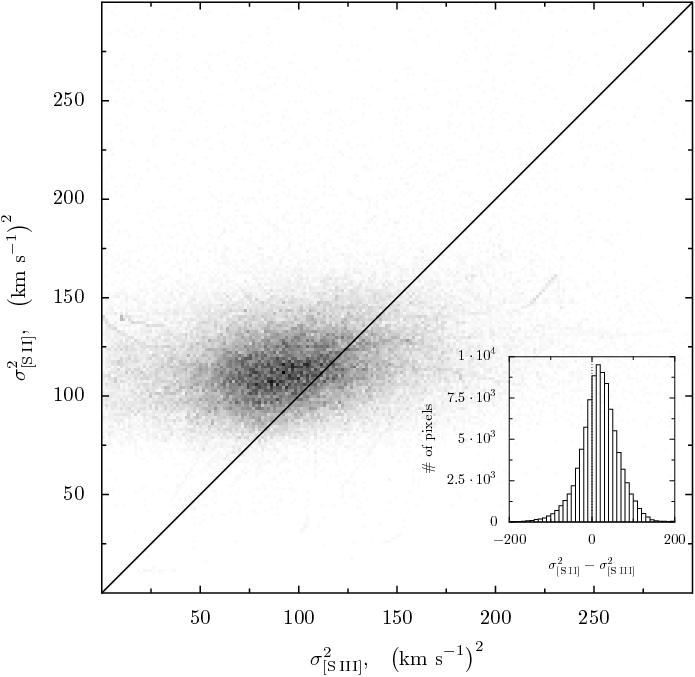}}\hfill
    \parbox[t]{\panelwidth}{\includegraphics{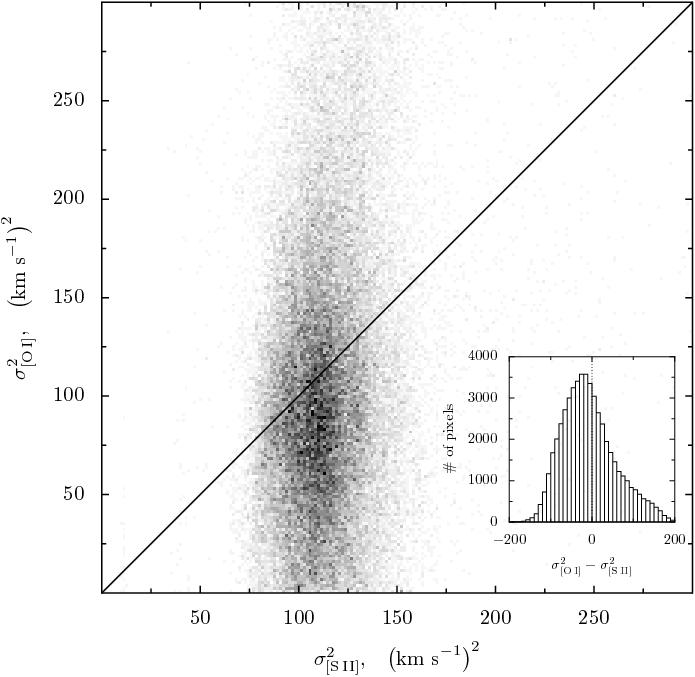}}
  \end{wide}
  \caption{Joint distribution of raw measured mean square velocity
    widths, $\sigma^2$, of different emission lines, with inset box
    showing a histogram of the differences in $\sigma^2$. (\textit{a})
    $\sigma^2_{\Ha}$ versus $\sigma^2_{\plainOIII}$.  (\textit{b})
    $\sigma^2_{\plainNII}$ versus
    $\sigma^2_{\plainOIII}$. (\textit{c}) $\sigma^2_{\plainSII}$
    versus $\sigma^2_{\plainSIII}$. (\textit{d}) $\sigma^2_{\plainOI}$
    versus $\sigma^2_{\plainSII}$.  The diagonal straight line in each
    panel shows the case of equal variances in the two lines, except
    for in panel \textit{a}, where it shows the relation
    $\sigma^2_{\Ha} = \sigma^2_{\plainOIII} + 82.2$. }
  \label{fig:sigmavssigma}
\end{figure*}

\begin{figure*}
  \begin{wide}
    \centering
    \setlength\panelwidth{0.48\linewidth}
    \parbox[t]{\panelwidth}{(\textit{a})}\hfill
    \parbox[t]{\panelwidth}{(\textit{b})}\\
    \parbox[t]{\panelwidth}{\includegraphics{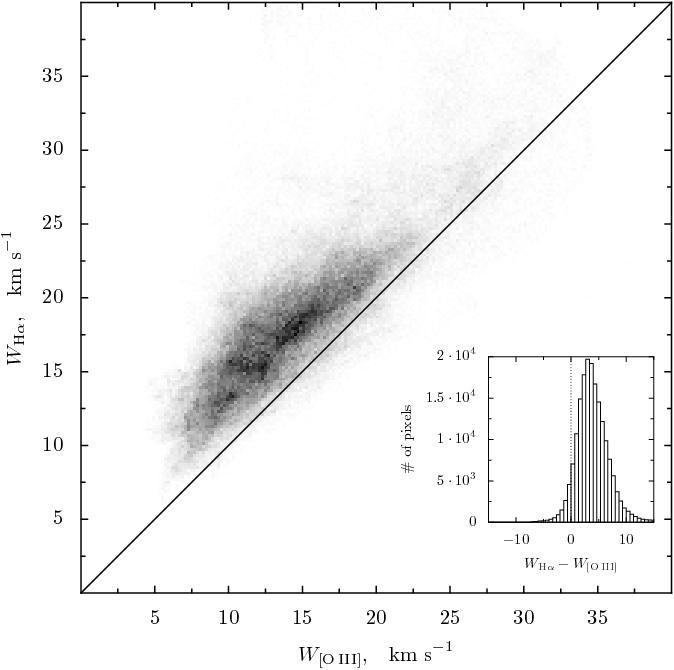}}\hfill
    \parbox[t]{\panelwidth}{\includegraphics{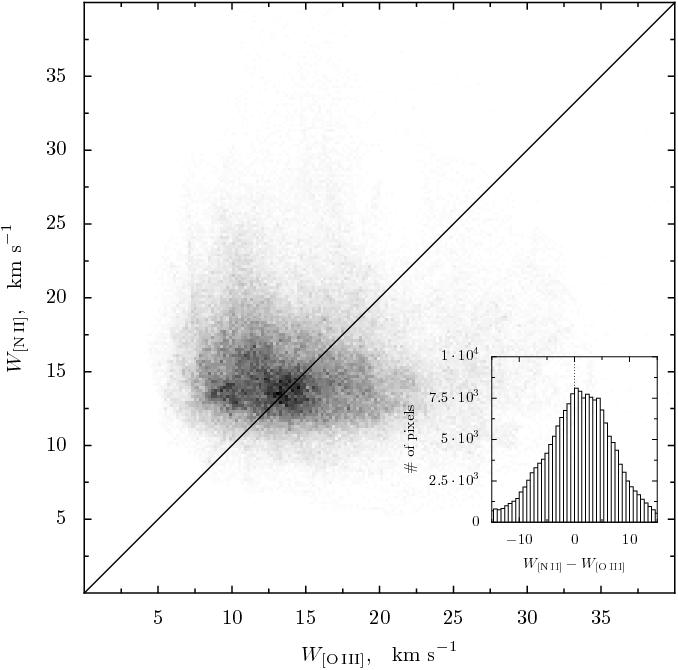}}
    \parbox[t]{\panelwidth}{(\textit{c})}\hfill
    \parbox[t]{\panelwidth}{(\textit{d})}\\
    \parbox[t]{\panelwidth}{\includegraphics{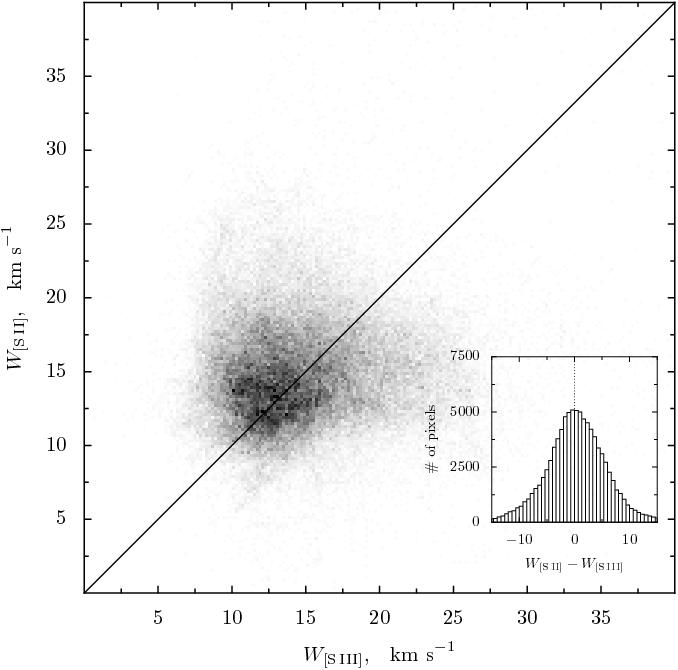}}\hfill
    \parbox[t]{\panelwidth}{\includegraphics{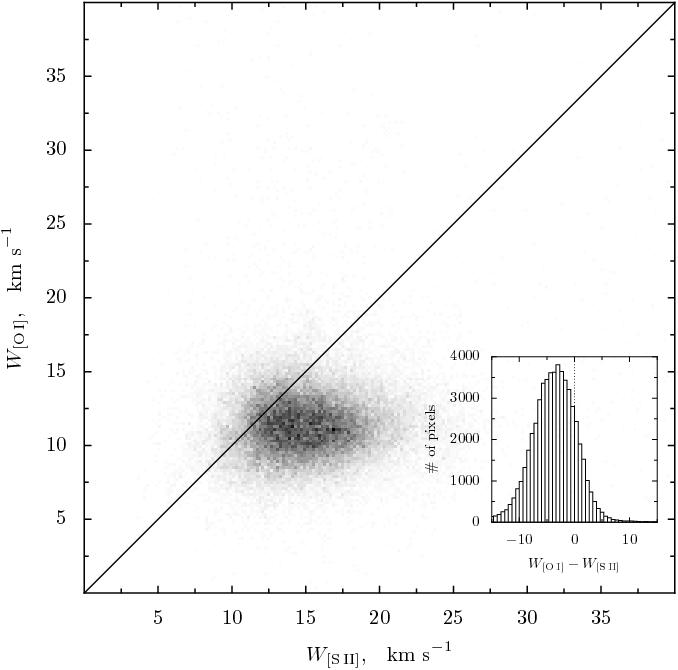}}
  \end{wide}
  \caption{Joint distribution of corrected full-widths at half
    maximum, $W$, of different emission lines, with inset box showing
    a histogram of differences in $W$. The thermal and instrumental
    widths have been subtracted in quadrature from the raw measured
    values. In each case, the straight line shows the case of equal
    widths in the two lines. (\textit{a}) $W_{\Ha}$ versus
    $W_{\plainOIII}$. (\textit{b}) $W_{\plainNII}$ versus
    $W_{\plainOIII}$ (\textit{c}) $W_{\plainSII}$ versus
    $W_{\plainSIII}$. (\textit{d}) $W_{\plainOI}$ versus
    $W_{\plainSII}$}
  \label{fig:fwhmvsfwhm}
\end{figure*}

In Figures~\ref{fig:meanvsmean} and \ref{fig:peakvspeak} we show joint
distributions of the mean and peak velocities, respectively, for
selected pairs of emission lines, which illustrate the correlations
that exist between the kinematics of the different emission zones of
the nebula. Figure~\ref{fig:sigmavssigma} shows the joint
distributions of the raw velocity variance, $\sigma^2$, which is the
square of the RMS velocity dispersion, and Figure~\ref{fig:fwhmvsfwhm}
shows distributions of the non-thermal FWHM linewidth, $W$. The values
of $W$ have been approximately corrected for thermal, instrumental and
fine-structure broadening in the same way as in
Figure~\ref{fig:moments-fwhm}, whereas the values of $\sigma^2$ in
Figure~\ref{fig:sigmavssigma} are uncorrected and measured directly
from the data. 

\newcommand\sd{s.d.\@} The statistics for individual lines are
summarised in the upper part of Table~\ref{tab:stats}, which shows the
mean and standard deviation (\sd{}) of each quantity. These are
calculated with an equal weighting of each pixel in the map after
masking out any points lying outside the ranges shown in
Figures~\ref{fig:meanvsmean}--\ref{fig:fwhmvsfwhm}. For \SIII{} and
\OI{}, any pixel with surface brightness below the thresholds shown in
Figures~\ref{fig:moments-mean}--\ref{fig:moments-fwhm} were also
masked out. The lower part of Table~\ref{tab:stats} show the mean and
\sd{} of the differences between the kinematic quantities for the line
pairs shown in the figures, together with the linear correlation
coefficient, $r$, of the joint distributions. In cases where the
quantities are well correlated ($r \gtrsim 0.5$), the \sd{} of the
difference can be significantly less than the \sd{}'s of the
individual values.

The general trend of increasing blueshifts with increasing degree of
ionization is readily apparent from the table, which is arranged from
highest to lowest ionization.  The mean and peak velocities always
show a positive correlation between different lines. The correlation
is highest ($r \simeq 0.9$) between H$\alpha$ and \OIII{}, which is
not surprising since there is a large overlap of the emission zones of
these two lines. However, even pairs of lines that should not greatly
overlap (such as $\NII{} - \OIII{}$ and $\SII{} - \SIII{}$) also show
significant correlation ($r > 0.5$). 

The \sd{} of 1.4~\kms{} for $\vmean_{\Ha} - \vmean_{\plainOIII}$ is a
hard upper limit on the uncorrelated average wavelength calibration
errors within the individual lines. Given that $\vmean_{\plainNII} -
\vmean_{\plainOIII}$ has a much larger \sd{} (2.8~\kms{}) and that
some fraction of the \Ha{} emission comes from the \NII{} zone, it
follows that at least some of the scatter in $\vmean_{\Ha} -
\vmean_{\plainOIII}$ must be real. This is consistent with our
estimate above in \S~\ref{sec:atlas} that the relative velocity
calibration within each emission line is good to $\simeq
1~\kms{}$. Our worst-case estimate for the uncertainty in the
\emph{absolute} velocity calibration for each line is 2~\kms{} and
this translates into uncertainties in the average values of the
differences in mean and peak velocities. This issue is returned to in
\S~\ref{sec:deriv-mean-electr} below, where we argue that the absolute
velocity calibration must be better than this conservative estimate.

An extremely tight correlation ($r \simeq 0.9$) is seen between the
mean square velocity widths of H$\alpha$ and \OIII{}, whereas no other
pair of lines show a significant correlation in this quantity. The
large offset $\bigl\langle \sigma^2_{\Ha} - \sigma^2_{\plainOIII}
\bigr\rangle = (82.2 \pm 18.2)~\kmssq$ is due mainly to the extra
thermal broadening of \Ha{}, as is discussed further in
\S~\ref{sec:deriv-mean-electr} below. Other pairs of lines show much
smaller offsets in $\sigma^2$, with no obvious trends with
ionization. The \sd{} of $\sigma^2$ is very large in the case of \OI{}
and \SIII{} because of the low signal-noise ratio of these weak lines.

A similar picture is seen for the corrected full width half maxima,
$W$: only in the case of $\Ha{} - \OIII{}$ does a correlation
exist. Note that even though the effects of thermal and fine structure
broadening have been subtracted in quadrature to obtain $W$, a
residual difference remains between \Ha{} and \OIII{} of $\simeq
4~\kms$. We believe that this is because the FWHM is such a
mathematically ill-behaved quantity that quadrature subtraction cannot
be expected to work reliably in the presence of strongly non-Gaussian
lineshapes. For example, consider a line that consists of two
components, A and B, separated by $\delta v$ and each of width $w$,
where the intensity of B is less than half that of A\@. The measured
FWHM will not be affected at all by the presence of component B unless
the components are blended: $\delta v \lesssim w$. The larger thermal
broadening of \Ha{} gives a larger $w$, so that there will be cases
where the FWHM of \Ha{} can ``see'' component B, but the FWHM of
\OIII{} cannot. Evidence for this can be seen in
Figure~\ref{fig:moments-fwhm}, where the \OIII{} map shows multiple
dark patches of low FWHM, which are absent from the \Ha{} map.

All the metal lines show very similar mean values of $\langle W\rangle
\simeq 15~\kms{}$, except for \OI{}, which is significantly lower,
$\langle W\rangle \simeq 12~\kms{}$. The FWHM is affected by noise to
a much smaller extent than $\sigma^2$ and the metal lines all have low
thermal widths, so should not be affected by the problem discussed in
the previous paragraph. 


\section{Derivation of the mean electron temperature from linewidths}
\label{sec:deriv-mean-electr}

In this section, we use the statistics of the observed line profiles
of \NII{}, \Ha{}, and \OIII{} (\S~\ref{sec:velocity-statistics}) to
derive the mean electron temperature in the nebula and to show that
the ``non-thermal'' component to the line broadening is not
significantly larger in \Ha{} than in the metal lines. The method we
employ is essentially an old one \citep{1968AnAp...31..493C,
  1971A&A....12..219D}, which estimates the gas temperature by
comparing the width of a metal line with that of a hydrogen line,
under the assumption that the excess width of the hydrogen line is
principally due to its greater thermal broadening on account of its
lower atomic weight. A problem with many applications of this method
has been that the metal line and hydrogen line (usually \NII{} and
\Ha{}) are not necessarily emitted by the same volume of gas. Indeed,
previous applications of this method to the Orion nebula
\citep{1973A&A....22...33D, 1976MNRAS.174..105G} have shown that the
\Hb{}-\OIII{} pair gives results that are very different from the
\Ha{}-\NII{} pair and that are seemingly more reliable. Our own
version of the method improves on previous versions by using all three
emission lines \NII{}, \Ha{}, and \OIII{} in order to control for the
imperfect overlap of the emission regions.

\subsection{Components of the observed line widths}
\label{sec:comp-observ-line}
For each emission line, the observed mean square velocity width at a
certain point on the nebula can be broken down schematically into four
components:
\begin{equation}
  \label{eq:sigsq}
  \sigma^2 = \sigma^2_\mathrm{th} + \sigma^2_\mathrm{fs}
  + \sigma^2_\mathrm{ins} + \sigma^2_\mathrm{nt} . 
\end{equation}
The first term in this equation is the thermal Doppler width,
$\sigma^2_\mathrm{th} = 82.5 (T_4/A)$ \kmssq{}, where $T_4 =
T/10^4~\mathrm{K}$ and $A$ is the atomic weight of the atom or
ion. The second term is the fine structure broadening, which is
important for hydrogen and helium recombination lines, but not for
metal lines such as \OIII{} and \NII{}. In
Appendix~\ref{sec:fine-struct-comp}, we show that its value for \Ha{}
is $\sigma^2_\mathrm{fs} = 10.233~\kmssq$. The third term is the
instrumental width, which for the KPNO observations is approximately $
\sigma^2_\mathrm{ins} = 11.5~\kmssq$, and which is the same for the
three lines. The fourth term, $\sigma^2_\mathrm{nt}$, is the
``non-thermal'' contribution, which is a catch-all term for any other
broadening process. This might include Doppler broadening due to
velocity gradients or turbulence within the emitting regions,
broadening due to scattering by dust particles, or any other process.

Applying the above equation to differences in $\sigma^2$ measured in
\S~\ref{sec:velocity-statistics} gives
\begin{multline}
  \label{eq:sigsqdiff1}
  \sigma^2_{\Ha} - \sigma^2_{\plainOIII} = 77.34 T_4 + 10.233\\ 
  + \sigma^2_\mathrm{nt,\Ha} - \sigma^2_\mathrm{nt,\plainOIII}
\end{multline}
and
\begin{equation}
  \label{eq:sigsqdiff2}
  \sigma^2_{\plainNII} - \sigma^2_{\plainOIII} = 0.74 T_4 
  + \sigma^2_\mathrm{nt,\plainNII} - \sigma^2_\mathrm{nt,\plainOIII} .
\end{equation}
Note that the instrumental contribution cancels from these equations.
The simplest way of extracting the temperature, and what has been done
in the past, is then to assume that the difference in non-thermal
widths is negligible, so that $T_4$ can be calculated directly from
the measured $\sigma^2_{\Ha} - \sigma^2_{\plainOIII}$ as
\begin{equation}
  \label{eq:Tnaive}
  T'_4 =  \bigl(\sigma^2_{\Ha} - \sigma^2_{\plainOIII} - 10.233\bigr)/77.34
\end{equation}
However, by including the data on the \NII{} line, it is possible to
derive a lower bound on the value of $\sigma^2_\mathrm{nt,\Ha} -
\sigma^2_\mathrm{nt,\plainOIII}$ and so arrive at a more reliable
temperature determination than the naive estimate $T'_4$.

\subsection{Correction for differences in temperature and non-thermal
  widths between high and low ionization zones}
\label{sec:corr-diff-non} 

We adopt a simplified two-zone model for the emission structure along
a given line of sight through the nebula. The first zone, A,
contributes all the \OIII{} emission plus a fraction, $f$, of the
\Ha{} emission. The second zone, B, contributes all the \NII{}
emission plus the remaining fraction, $1-f$, of the \Ha{} emission. We
further assume that the non-thermal broadening is the same for all
emission lines that arise from the same volume (this assumption should
hold true for turbulent or other kinematic broadening, an issue we
will return to later). A fundamental property of the moments of the
line profile given in equation~(\ref{eq:moments}) is that they are all
\emph{additive}, with consequences that are elucidated in
Appendix~\ref{sec:prop-veloc-moments}. From
equation~\eqref{eq:sigcombo} it follows that
\begin{multline}
  \label{eq:twozone}
  \sigma^2_\mathrm{nt,\Ha} - \sigma^2_\mathrm{nt,\plainOIII} = 
  (1 - f) (\sigma^2_\mathrm{nt,\plainNII} -
  \sigma^2_\mathrm{nt,\plainOIII})\\
  + f (1 - f) \bigl( \vmean_{\plainNII} - \vmean_{\plainOIII}
  \bigr)^2 .
\end{multline}
Note that the above equation is only valid if the effects of dust
extinction are the same for zones A and B, but this is probably a good
approximation since \citet{2000AJ....120..382O} showed that the
majority of the dust extinction in Orion arises in the neutral veil,
in the foreground of the ionized gas.  

We also allow the two emission zones to have different temperatures,
$T_{\mathrm{A}}$ and $T_{\mathrm{B}}$, such that the mean \Ha{}
temperature is $T = f T_{\mathrm{A}} + (1-f) T_{\mathrm{B}}$.  We can
now combine equation~(\ref{eq:twozone}) with
equations~(\ref{eq:sigsqdiff1}) and~(\ref{eq:sigsqdiff2}), modified to
account for the temperature difference between the zones, to obtain
\begin{equation}
  \label{eq:t4}
  T_4 = 
  \eta T'_4 + \epsilon - \zeta - \xi , 
\end{equation}
where
\begin{gather}
  \label{eq:eta}
  \eta = \bigl(1 - 0.00957\, (1 - f)\bigr)^{-1}, \\
  \label{eq:epsilon}
  \epsilon = 0.00957\, \eta\, f (1 - f) \bigl( T_{\mathrm{B}} - T_{\mathrm{A}} \bigr)/
  10^4~\mathrm{K}, \\
  \label{eq:zeta}
  \zeta = 0.01293\, \eta\, (1 - f) \bigl(\sigma^2_{\plainNII} - \sigma^2_{\plainOIII} \bigr), \\
  \label{eq:xi}
  \xi = 0.01293\, \eta\, f (1 - f) \bigl(\vmean_{\plainNII} -
    \vmean_{\plainOIII}\bigr)^2 . 
\end{gather}

\subsection{Techniques for estimating the ionization factor~\lowercase{\(f\)}}
\label{sec:techn-estim-fact}
Equation~(\ref{eq:t4}) gives the temperature at any point in the
nebula in terms of observable quantities and the factor $f$, which can
be estimated in two ways: from surface brightness ratios, or from
ratios of differences in mean velocity. We shall describe each of
these techniques in turn. 

We define for each emission line a per-zone mean effective emission
coefficient, $\emco$, with units $\mathrm{erg\ cm^3\ s^{-1}\
  sr^{-1}}$, such that the intrinsic surface brightness of a zone is
$S = \emco \EM$, where $\EM = \int n_\mathrm{p} n_\mathrm{e}\,dz$ is
the emission measure of the zone.  Since \NIIlam{} and \Halam{} are so
close in wavelength, they are both affected in the same way by dust
extinction, so the observed ratio of their surface brightnesses is
equal to the intrinsic value. Therefore, given that \NII{} by
construction arises only in Zone~B, we have
\begin{multline}
  \label{eq:sratio}
  R_{\plainNII} \equiv \frac{S(\NII)}{S(\Ha)} = \frac{ \emco\B(\NII)\,
    \EM\B }{\emco\A(\Ha)\, 
    \EM\A + \emco\B(\Ha)\, \EM\B} \\ = (1 - f) \frac{\emco\B(\NII)
  }{\emco\B(\Ha)}, 
\end{multline}
where the last equality follows from the definition of $f$. Hence, if
the emission coefficients $\emco\B(\NII)$ and $\emco\B(\Ha)$ are
known, then $f$ can be calculated from the observed surface brightness
ratio as $f = 1 - R_{\plainNII} \emco\B(\Ha) / \emco\B(\NII)$. To make
use of this equation, it is first necessary to photometrically
calibrate the surface brightness maps and we have done this by
comparison with published spectrophotometry of the region
\citep{1992ApJ...399..147P, 2000ApJS..129..229B,
  2003AJ....125.2590O}. It is then necessary to estimate the ratio of
emission coefficients for conditions typical of Zone~B: $T \simeq
10^4$~K, $n_{\mathrm{e}} \simeq 5000$, which we have done using the
Cloudy plasma code \citep{2000RMxAC...9..153F}, assuming an N/H
abundance ratio of $6\times 10^{-5}$, finding that the ratio is
roughly unity. This is consistent with the N\p/N abundance found by
\citet{1998MNRAS.295..401E}.

\begin{figure*}
  \begin{wide}
    \centering
    \includegraphics{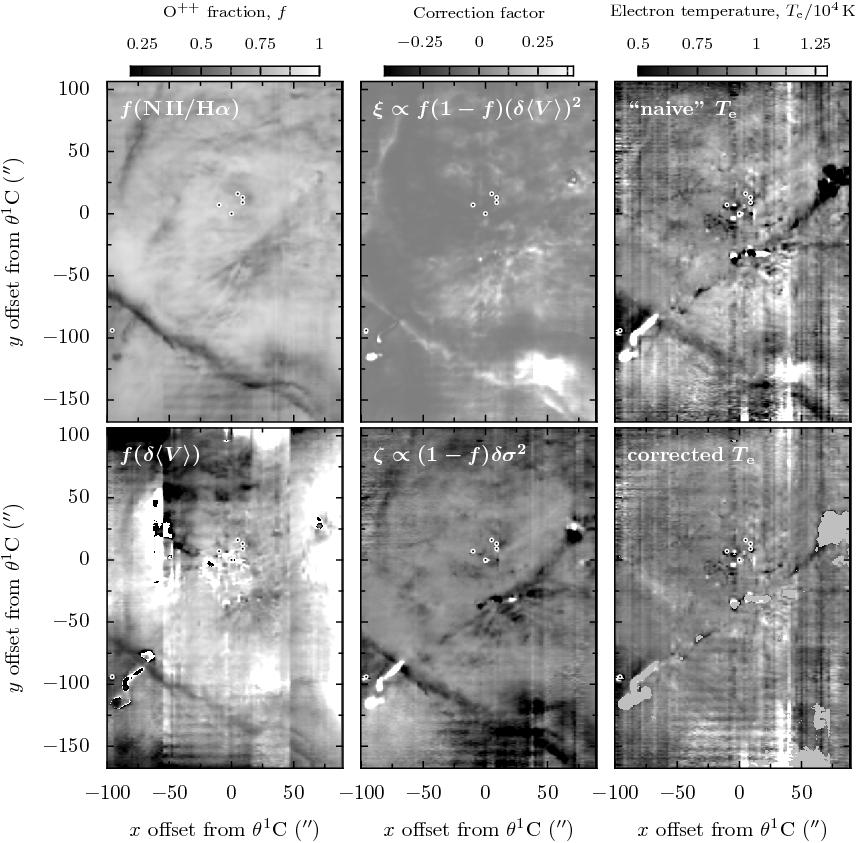}
  \end{wide}
  \caption{Steps in the construction of the temperature
    map. \textit{Left column:} Two independent estimates of the
    fraction, $f$, of the \Ha{} emission that comes from the O$^{++}$
    zone: from the $\NII/\Ha$ surface brightness ratio (upper panel),
    and from the differences in mean velocities of \NII{}, \Ha{},
    \OIII{} (lower panel). \textit{Center column:} Correction factors
    to the ``naive'' temperature map due to differences in the
    kinematics between the N$^{+}$ and O$^{++}$ zones: mean velocity
    differences, $\xi$ (upper panel) and velocity dispersion
    differences, $\zeta$ (lower panel). \textit{Right column:}
    ``naive'' temperature estimate (upper panel) from the difference
    in velocity dispersion of \OIII{} and \Ha{} and final corrected
    temperature estimate (lower panel). The horizontal bands and
    vertical streaks seen in some panels are instrumental artifacts.}
  \label{fig:tmaps}
\end{figure*}
Assuming that $\emco\B(\Ha) / \emco\B(\NII) = 1$, we find a mean value
of $\langle f \rangle = 0.76$ for the entire area covered by our
observations. The upper-left panel of Figure~\ref{fig:tmaps} shows the
map of $f$ calculated by this technique, which varies between 0.26 and
0.91. In reality, $\emco\B(\Ha) / \emco\B(\NII)$ will vary with
position due to variations in electron density (\NII{} is affected by
collisional deexcitation but \Ha{} is not). A better way of
calculating $f$ would be to use the $\OIII/\Hb$ surface brightness
ratio, since the critical density for \OIII{} is higher than that
typically found in the nebula (similar arguments to those above give
$f = R_{\plainOIII} \emco\A(\Hb) / \emco\A(\OIII)$, where
$R_{\plainOIII} = S(\OIII)/S(\Hb)$). However, our dataset does not
include \Hb{} and the $\OIII/\Ha$ ratio would have to be corrected for
the highly variable foreground extinction
\citep{2000AJ....120..382O}. A comparison of the $\NII/\Ha$ and
$\OIII/\Hb$ ratios over a limited area (using data from Figs.~11 and
12 of \citealp{2003AJ....125.2590O}) gives an rms scatter of $\pm0.05$
between the values of $f$ derived from the $R_{\plainNII}$ and
$R_{\plainOIII}$ techniques, which is roughly half of the total rms
variation in $f$ from our maps. The $R_{\plainOIII}$-derived values of
$f$ have a greater range than the $R_{\plainNII}$-derived values. This
is to be expected since, all else being equal, higher density regions
will have lower ionization.

\begin{figure}
  \centering
  \includegraphics{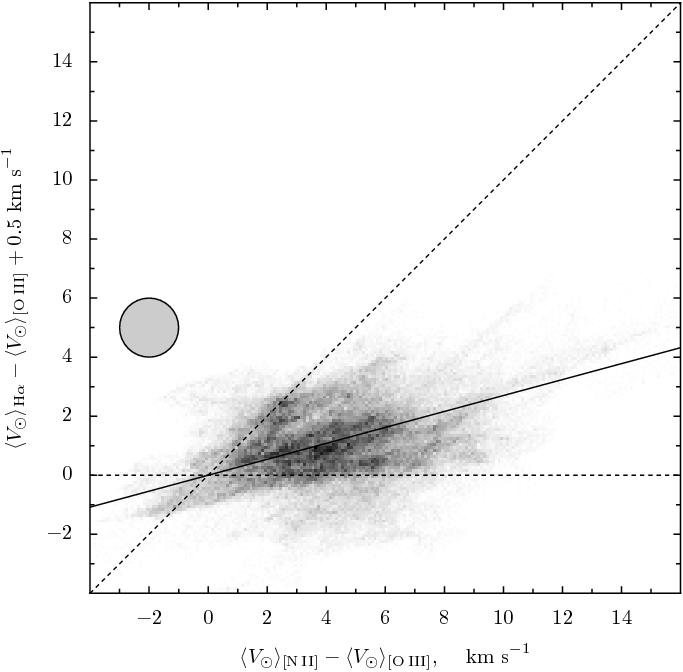}
  \caption{Joint distribution of differences in mean velocities:
    $\vmean_{\Ha} - \vmean_{\plainOIII}$ versus $\vmean_{\plainNII} -
    \vmean_{\plainOIII}$. The values of $\vmean_{\Ha} -
    \vmean_{\plainOIII}$ have been shifted by 0.5~\kms{} so as to make
    the distribution symmetric about a line through the origin (see
    discusion in text). The dashed lines show the boundaries of the
    two triangular regions (upper right and lower left) that give
    physically possible values of $f$ in the two-zone model. The solid
    line shows the relationship $\vmean_{\Ha} - \vmean_{\plainOIII} =
    (1 - \langle f\rangle)(\vmean_{\plainNII} - \vmean_{\plainOIII})$,
    where $\langle f\rangle$ is determined from the $R_{\plainNII}$
    method. The solid gray circle indicates the effect of $\pm 1~\kms$
    uncertainties in the velocities.  }
  \label{fig:vmean-diff}
\end{figure}
A completely independent way of estimating $f$, and one that does not
depend on knowing the emission coefficients, is to use the observed
mean velocities of the three lines. A corollary of the additive
property of the line profile velocity moments
(Appendix~\ref{sec:prop-veloc-moments}) is that in the two-zone model
we have from equation~\eqref{eq:meancombo} that
\begin{equation}
  \label{eq:addvmean}
  \vmean_{\Ha} = f \vmean_{\plainOIII} + (1 - f) \vmean_{\plainNII},
\end{equation}
so that $f$ can be estimated as
\begin{equation}
  \label{eq:fvmean}
  f = 1 - \frac{\vmean_{\Ha} - \vmean_{\plainOIII}}
  {\vmean_{\plainNII} - \vmean_{\plainOIII}}. 
\end{equation}
Despite its advantages over the surface brightness method, this
technique has two serious limitations. First,
equation~(\ref{eq:fvmean}) has a singularity when $\vmean_{\plainNII}
= \vmean_{\plainOIII}$. If the two-zone model were strictly valid,
then any pixel with $\vmean_{\plainNII} = \vmean_{\plainOIII}$ would
also have $\vmean_{\Ha} = \vmean_{\plainOIII}$, so the singularity
would be benign. However, in reality the two-zone model is only an
approximation and, additionally, the observed values of \vmean{} will
always include a certain measurement error. As a result, applying
equation~(\ref{eq:fvmean}) to real data will inevitably result in
discontinuities in the $f$ map and unphysical values ($f <0$ or $f >
1$) for some pixels. Second, the observed values of $\vmean_{\Ha}
-\vmean_{\plainOIII}$ are rather small (typically $-2$ to $+3$~\kms{},
see Fig.~\ref{fig:meanvsmean}\textit{a}), and are of the same order as
the systematic uncertainties in the wavelength calibration. 

Given these caveats, it is perhaps surprising that this method works
at all. However, the average \emph{slope} of the relationship between
$\vmean_{\Ha} - \vmean_{\plainOIII}$ and $\vmean_{\plainNII} -
\vmean_{\plainOIII}$ should be independent of any inaccuracies in the
absolute velocity scales of the individual lines, and by
equation~(\ref{eq:fvmean}) should be equal to $1 - \langle
f\rangle$. This is illustrated in Figure~\ref{fig:vmean-diff}, where
it can be seen that the value $\langle f\rangle = 0.76$, calculated
from the $R_{\plainNII}$ method above and shown by the solid line, is
indeed close to the mean trend in the velocity
differences. Furthermore, to the extent that the two-zone
approximation is valid, the symmetry axis of the joint distribution of
velocity differences ought to pass through the origin. This is only
true when we shift all our $\vmean_{\Ha} - \vmean_{\plainOIII}$ values
by $+0.5~\kms{}$ and this is what is done in the figure. Such a small
shift is well within the estimated uncertainties in the velocity zero
point of each line. Of course, there may also be a similarly small
constant error in the $\vmean_{\plainNII} - \vmean_{\plainOIII}$
values, but we have no information that would allow us to determine
the correction in this case.

In order to correspond to a value of $f$ between 0 and 1, the velocity
differences must lie in one of the two triangular areas of the graph
in Figure~\ref{fig:vmean-diff}, bounded by dashed lines. It can be
seen from the graph that the majority of pixels do indeed lie in these
areas, mostly on the positive side of the origin. The small fraction
of pixels lying in unphysical areas of the graph can be explained by
residual errors in the velocity determination of order $1~\kms$
(indicated by the solid gray disk). The lower left panel of
Figure~\ref{fig:tmaps} shows the map of $f$ calculated using this
velocity difference technique. For much of the map, the agreement with
the surface brightness ratio technique (upper left panel) is rather
good, and the low ionization regions associated with the linear bright
bars are readily apparent in both maps. The unphysical values of $f$
show up as solid black or white regions in the map. The fact that
these regions occur predominantly at the edges of the map, with white
regions ($f >1$) concentrated in the West and black regions ($f < 0$)
in the East, suggests that the residual velocity errors are largely
systematic, rather than random.

\subsection{Corrected value of the electron temperature}
\label{sec:corr-value-electr}
\begin{table}
  \centering
  \caption{Summary of temperature determinations}
  \label{tab:temp}
  \begin{tabular}{rrrrr}
    \toprule
    &  \(\langle T_4 \rangle\) & \(\langle T_4 \rangle\avS\) 
    &  $t^2_{\mathrm{A}}$ &      $N_{\mathrm{pix}}$ \\
    \midrule
    naive            & 0.930 &  0.942 & 0.025 & 178432 \\
    corrected        & 0.913 &  0.916 & 0.014 & 178432 \\
    low \(\sigma\)   & 0.916 &  \textbf{0.919} & \textbf{0.012} & 173180 \\
    high \(\sigma\)  & 0.824 &  0.792 & 0.104 &   5252 \\
    \bottomrule    
  \end{tabular}
\end{table}
We are finally in a position to calculate the correction factors,
$\epsilon$, $\zeta$, and \(\xi\) needed to estimate the electron
temperature. The mean values are $\langle\epsilon\rangle = 2.9 \times
10^{-4}$, \(\langle\zeta\rangle = -0.0218\) and \(\langle\xi\rangle =
0.0417\), so that the corrected mean temperature from
equation~(\ref{eq:t4}) is \(\langle T_4 \rangle = 0.913\), whereas the
naive estimate would have been \(\langle T'_4 \rangle = 0.930\) (note
that all these averages are calculated with equal weighting for each
pixel). Maps of the spatial variation of the correction factors
\(\zeta\) and \(\xi\) are shown in the central panels of
Figure~\ref{fig:tmaps}, while the right-hand panels show maps of the
naive and corrected temperatures. The correction factor \(\epsilon\)
is uniformly very small and will not be considered further. The effect
of the corrections on the mean temperature is rather small, merely
lowering it by 200~K, but, as can be seen from the maps, the
correction factors can be much larger in localized regions (\(>
2000~\mathrm{K}\)). Some of the variation seen in the uncorrected
\(T'_4\) map is removed by the correction factors, with the most
notable example being at the bright bar, which is seen as a dark
diagonal band in the \(T'_4\) map, but not at all in the \(T_4\) map,
principally due to \(\zeta\). Another example is the ``Red Fan''
\citep{2007AJ....133..952G}, seen as a light feature in the lower
right of the \(T'_4\) map, but which is effectively removed by the \(\xi\)
correction (the thin vertical streaks seen superposed on this feature
are observational artifacts, which can also be seen in the
$\sigma_{\Ha}$ map of Fig.~\ref{fig:moments-sigma}). Other striking
features seen in the temperature maps are those associated with the
high-velocity jets emanating from the Orion~South region
\citep{2007AJ....133.2192H}: \HH{202} (northeast), \HH{203/204}
(southwest), and \HH{529/269} (east-west). These features also tend to
have high absolute values of \(\zeta\) or \(\xi\), but the application
of the correction factors is not successful in removing them from the
temperature map. Despite this, we do not believe that they correspond
to real temperature variations. Since all these features have large
rms velocity widths (Fig.~\ref{fig:moments-sigma}), we have masked out
of the corrected map all pixels with $\sigma^2_{\mathrm{\plainOIII}}$
or $\sigma^2_{\mathrm{\plainNII}}$ greater than 180~\kmssq{} (visible
as small solid gray areas in the figure).

In order to quantify the variation in the naive and corrected
temperature maps, we calculate the plane-of-sky Peimbert $t^2$
parameter \citep{1967ApJ...150..825P, 2003AJ....125.2590O}:
$t_{\mathrm{A}}^2 = \bigl\langle (T - \langle T\rangle\avE)^2
\bigr\rangle\avE / \langle T\rangle\avE$. The averages in this
expression, $\langle\cdots\rangle\avE$, should theoretically be
weighted by the emission measure of each pixel, but we approximate
this by weighting by the \Ha{} surface brightness:
$\langle\cdots\rangle\avS$. The results are summarised in
Table~\ref{tab:temp}. It can be seen that the \Ha{}-weighted averages,
\(\langle T_4 \rangle\avS\), are hardly any different from the
uniform-weighted averages, \(\langle T_4 \rangle\). The fractional
variance, \(t^2_{\mathrm{A}}\), is significantly reduced after
applying the temperature corrections (second row of table). A further
small reduction in \(t^2_{\mathrm{A}}\) occurs (third table row) when
the high-\(\sigma\) pixels are eliminated as discussed above. The
high-\(\sigma\) pixels themselves (fourth table row) show a
significantly lower mean temperature and a very large
\(t^2_{\mathrm{A}}\).

\section{Discussion}
\label{sec:discussion}

\subsection{Comparison with previous mean temperature determinations}
\label{sec:comp-with-prev}
Previous determinations of the mean value of the electron temperature
in the Orion Nebula have been made using various methods (see
\citealp{2006agna.book.....O}, Chapter~5 for an overview). In this
section, we compare our own results (\S~\ref{sec:corr-value-electr})
with some recent examples of these.

\citet{2003AJ....125.2590O} used \textit{HST} emission line imaging in
the \OIII{} 4363~\AA{} and 5007~\AA{} lines to determine
$T_{\plainOIII} = 8890 \pm 496$~K, where the scatter represents real
variations, rather than observational uncertainties. Before comparing
with our own result, this temperature needs to be corrected in two
ways. First, temperature fluctuations \emph{within} the
\OIII{}-emitting region will lead to this method overestimating the
real temperature there, $T(\mathrm{O}^{++})$. Second,
$T(\mathrm{O}^{++})$ (which corresponds to \(T_\mathrm{A}\) from
\S~\ref{sec:corr-diff-non} above) is an underestimate of the mean
temperature of the ionized gas, $T(\mathrm{H}^{+})$, since it does not
include the low ionization zone along each line of sight (Zone~B),
which tends to have a higher temperature. Taking estimates of the
magnitudes of these corrections from \citet{1998MNRAS.295..401E} and
\citet{2003MNRAS.340..362R}, we find that the
\citet{2003AJ....125.2590O} results imply $T(\mathrm{O}^{++}) =
8400$~K and $T(\mathrm{H}^{+}) = 9100$~K, which is very similar to our
result of 9190~K (Table~\ref{tab:temp}).

Electron temperatures have also been derived from the brightness
relative to the continuum of radio recombination lines
\citep{1984A&A...138..225W, 1987A&A...184..291W, 1997A&A...327.1177W},
consistently finding values of around $8500$~K\@. Yet another
technique is to measure the strength of the Balmer discontinuity at
3646~\AA{}, which was used by \citet{1995ApJ...450L..59L} to measure a
mean value of $T \simeq 9000$~K, whereas \citet{1998MNRAS.295..401E}
used the same technique to find $T \simeq 8600$~K\@. In the presence
of temperature fluctuations, these values are underestimates and
should be increased by $\simeq 300$~K\@. However, the experimental
uncertainties of these results are probably higher than for the other
methods and furthermore they only correspond to restricted regions of
the nebula.

In summary, once temperature fluctuations are accounted for, there is
a very good agreement between the three independent optical
determinations of the mean electron temperature (line ratios, Balmer
discontinuity, and line widths), giving an average value of $\simeq
9100 \pm 100$~K\@. The radio-derived temperatures are significantly
lower by about $600$~K and we have no satisfactory explanation for
this discrepancy.

\subsection{Plane-of-sky variation in the electron temperature}
\label{sec:plane-sky-variation}

The point-to-point variations in electron temperature across the face
of the nebula have been studied in detail by two recent papers
\citep{2003AJ....125.2590O, 2003MNRAS.340..362R} based on optical line
ratios from \textit{HST} observations. The values obtained for
$t^2_{\mathrm{A}}(\mathrm{O\pp})$ for different regions of the nebula
range from 0.005 to 0.016. \citeauthor{2003MNRAS.340..362R} also
calculate $t^2_{\mathrm{A}}(\mathrm{N\pp})$, finding broadly similar
values. After correcting for the effects of noise,
\citeauthor{2003AJ....125.2590O} find a value for their entire
surveyed region (roughly one third the area of our own dataset) of
$t^2_{\mathrm{A}}(\mathrm{O\pp}) = 0.0079$. Given that our own value
of \(t^2_{\mathrm{A}} = 0.012 \) will also be somewhat affected by
noise, the agreement may at first seem satisfactory. However, the
spatial resolution of our own observations is at least an order of
magnitude lower than these two studies, which find that power spectrum
of the fluctuations peaks at sub-arcsecond scales. Such fluctuations
would be entirely smoothed out by atmospheric seeing in our own
dataset, so one would expect our results to show a much lower value of
\(t^2_{\mathrm{A}}\). 

An analysis of the temperature variations along the slits used in the
\citeauthor{2003MNRAS.340..362R} study supports this point of view.
Their Slit~5 is the only one that does not cross high-velocity jet
features, and we find \(t^2_{\mathrm{A}} = 0.002\) for this slit,
which is very much lower than their value of 0.018, as is to be
expected from our lower spatial resolution. It therefore seems likely
that much of the variation visible in even our corrected temperature
map (lower right panel of Fig.~\ref{fig:tmaps}) is spurious, and that
\(t^2_{\mathrm{A}} = 0.002\) is a more accurate estimate of the
plane-of-sky temperature fluctuations at scales \(> 1\)~arcsec. This
is also consistent with calculations of the variation in three small
\(40\arcsec \times 40\arcsec\) patches of our map, which were selected
for their high signal-noise ratio and avoidance of high-velocity
flows. The resulting values are \(t^2_{\mathrm{A}} = 0.002\), 0.004,
and 0.002 for regions centered at \((-30,+60)\), \((-30,-100)\), and
\((-80,-50)\), respectively.

\subsection{The nature of the non-thermal line broadening}
\label{sec:the-nature-non}
A fundamental assumption of our method of determing the temperature is
that the non-thermal broadening depends only on the emitting volume
(see \S~\ref{sec:corr-diff-non}), and does not vary systematically
between collisionally excited and recombination lines. This is
necessary so that \(\sigma^2_{\mathrm{nt,\Ha}}\) can be determined
from the kinematics of the \NII{} and \OIII{} lines, as in
equation~(\ref{eq:twozone}). The fact that we determine a mean nebular
temperature that closely agrees with that from other techniques
(\S~\ref{sec:comp-with-prev}) is a strong argument that this
assumption is valid.

However, this is in conflict with the result of
\citet{2003AJ....125.2590O}, who found non-thermal widths (FWHM) of
\(\simeq 19~\kms\) for recombination lines (\Ha, \Hb, and He~I
5876~\AA), but only \(\simeq 11~\kms\) for collisional lines (\OI{},
\SII{}, \NII{}, \SIII{}, \OIII{}). Taken at face value, these widths
would imply a large difference of \((19^2 - 11^2) / (8 \ln 2) =
43~\kmssq\) between \(\sigma^2_{\mathrm{nt,\Ha}}\) and
\(\sigma^2_{\mathrm{nt,\plainOIII}}\), meaning that over half of the
difference \(\sigma^2_{\Ha} - \sigma^2_{\plainOIII}\)
(Fig.~\ref{fig:sigmavssigma}\textit{a}) is due to non-thermal effects.
In this case, the mean electron temperature would only be \(\sim
4000\)~K, in stark contrast to all other temperature determinations
for the nebula.
 
Instead, we suggest that the difference in non-thermal widths found by
\citeauthor{2003AJ....125.2590O} is an artifact of their methodology,
which consisted in fitting Gaussian components to the line shapes. The
basic problem with this is that the extra thermal broadening of the H
and He lines tends to blend the fine details of the line shape,
allowing a satisfactory fit to be obtained using fewer Gaussian
components than are necessary for the metal lines. These fewer
components would then naturally tend to have larger individual widths
(even after correcting for thermal broadening), which may explain the
\citeauthor{2003AJ....125.2590O} result. Our own method of calculating
the RMS linewidths does not suffer from this problem, since it avoids
any subjective decision about how many components to fit. Indeed, an
application of our methodology to the \emph{same data} as used by
\citeauthor{2003AJ....125.2590O} shows no evidence for extra
non-thermal broadening of the recombination lines (Fig.~4 of
\citealp{1999AJ....118.2350H}). A further problem with the
\citet{2003AJ....125.2590O} study is that no correction was made for
the H fine-structure broadening (although the larger He~I fine
structure was accounted for).

A third methodology was employed by \citet{1976MNRAS.174..105G}, who
convolved the observed \OIII{} profile with Gaussians of different
widths and determined which width gave the best match to the observed
\Hb{} profile. This technique gave temperatures that are very similar
to those derived here, which is further evidence that the
\citeauthor{2003AJ....125.2590O} widths cannot be correct.

One promising class of explanations for the non-thermal linewidths is
that they are kinematic in nature, caused by velocity structure in the
ionized gas. This may be due to velocity gradients in ordered,
large-scale champagne flows \citep[e.g.][]{1979A&A....71...59T,
  1984A&A...138..325Y, 2005ApJ...627..813H}, or to multi-scale,
turbulent motions \citep[e.g.][]{2006ApJ...647..397M,
  2007dmsf.book..103H}. However, as \citet{2005ApJ...627..813H} showed
in their \S~6.1.2, velocity gradients in an ordered flow are incapable
of explaining the observed linewidths. Furthermore, the disordered
appearance of the maps of the line profile parameters
(Figs.~\ref{fig:moments-mean} to \ref{fig:moments-fwhm}) would favor
the turbulent explanation. In the radiation-hydrodynamical simulations
of \citet{2006ApJ...647..397M}, the turbulence in the ionized gas is
driven primarily by photoevaporation flows from dense concentrations
of molecular gas, which are themselves the product of supersonic
turbulence in the molecular cloud from which the high-mass star
formed. The internal velocity dispersion of the \HII{} region remains
roughly sonic for the full duration of the simulations (\(\simeq
0.5\)~Myr), in agreement with the observed linewidths in Orion. In
addition, it is quite possible that other mechanisms may be
contributing to the turbulent driving, such as ionization front
instabilities \citep{1996ApJ...469..171G, 1999MNRAS.310..789W,
  2007astro.ph..3463W} or supersonic jets from young stars.

\section{Conclusions}
\label{sec:conclusions}

We have presented an atlas of spatially resolved high resolution
emission line spectra of the Orion Nebula, covering a wide range of
degrees of ionization. This has allowed us to investigate the global
kinematics of the nebular gas in unprecedented depth. We have studied
the statistical correlations between the line profile parameters of
the different emission lines. Our principal conclusions are as follows:

\begin{enumerate}
\item The mean velocities of different emission lines are always
  positively correlated. The correlation is greatest when there is
  considerable spatial overlap between the emitting volumes of the two
  lines, but correlation also exists between lines from
  non-overlapping volumes. This is indicative of the existence of an
  ordered mean flow that connects the spatially separate
  regions. 

\item There is no significant correlation, on the other hand, between
  the linewidths of emission lines from non-overlapping volumes, such
  as \NIIlam{} and \OIIIlam{}. This is consistent with kinematic
  broadening of an essentially turbulent nature. The kinetic energies
  associated with both ordered and unordered motions are of the same
  order as the thermal energy. 

\item The mean electron temperature in the nebula is determined to be
  9190~K from comparison of the widths of \Ha{} and \OIII{} lines,
  after correction for differences in non-thermal widths between the
  different emission zones. Although apparent spatial variations
  in temperature are detected (\(t^2_{\mathrm{A}} = 0.012\)), most are
  probably due to uncertainties in these corrections, rather than real
  temperature variations. In regions of the nebula where these
  correction terms are small, the spatial variations are much smaller
  (\(t^2_{\mathrm{A}} \simeq 0.002\)) at the scales accessible to our
  observations (\(> 1~\arcsec\)).

\item We find that there can be no systematic difference in the
  non-thermal linewidths between collisionally excited and
  recombination lines. This contradicts a recent claim that the
  recombination lines are significantly broader
  \citep{2003AJ....125.2590O}.

\end{enumerate}

\acknowledgments We acknowledge financial support from DGAPA-UNAM
projects PAPIIT IN112006 and IN110108, and IN116908. MTGD is supported
by a postdoctoral research grant from CONACyT, Mexico. We are
extremely grateful to Bob O'Dell for many discussions and advice, and
for freely sharing observational data. We thank Michael Richer, John
Meaburn, Alex Raga, and Jorge Garc\'\i{}a-Rojas for helpful
discussions. The referee is thanked for a useful report. 

\begin{appendices}

\section{Some properties of the velocity moments}
\label{sec:prop-veloc-moments}
\setcounter{equation}{0}
The velocity moments $M_k$ of an emission line, as defined in
equation~\eqref{eq:moments}, are all linear functions of the line
intensity $I(V)$. This means that when two or more optically thin
emission regions (A, B, C, \dots) are superimposed along the same line
of sight then the moments of the combined emission can be found by
simple addition of the moments of the individual regions:
\begin{gather}
  \label{eq:momsum}
  M_0 = M_{0,A} + M_{0,B} + M_{0,C} + \cdots , \\
  M_1 = M_{1,A} + M_{1,B} + M_{1,C} + \cdots , \\
  M_2 = M_{2,A} + M_{2,B} + M_{2,C} + \cdots .
\end{gather}

The first three formal moments $M_k$ are related to the more physical
quantities:
\begin{gather}
  \label{eq:intensity-def}
  \text{total intensity: } I = M_0 ,\\
  \label{eq:vmean-def}
  \text{mean velocity: } \vmean = M_1/M_0 ,\\
  \label{eq:sigma-def}
  \text{rms velocity width: } \sigma = \bigl(M_2/M_0 - \vmean^2\bigr)^{1/2} .
\end{gather}
These too can be calculated for the superposition of multiple
optically thin regions. In particular, for the superposition of two
regions, A and B, one can combine
equations~\eqref{eq:momsum}--\eqref{eq:sigma-def} to find
\begin{gather}
  \label{eq:intensitycombo}
  I = I_{A} + I_{B} ,\\
  \label{eq:meancombo}
  \vmean = \frac{\vmean_{A} I_{A} + \vmean_{B} I_{B}}{I_{A} + I_{B}} ,
  \\
  \label{eq:sigcombo}
  \sigma^2 = \frac{\sigma^2_{A} I_{A} + \sigma^2_{B} I_{B}}{I_{A} +
    I_{B}} + \frac{I_A I_B (\vmean_{A} - \vmean_{B})^2 }{(I_{A} +
    I_{B})^2} . 
\end{gather}

\section{Fine structure components of Hydrogen Balmer lines}
\label{sec:fine-struct-comp}

The fine structure components of the Balmer lines have velocity
differences of $< 7~\kms{}$, which are too small to cause observable
line splitting at typical nebula temperatures, where the Doppler
broadening is of order $20~\kms{}$. However, they do make a
significant contribution to the observed line width and, furthermore,
must be considered in order to make an accurate determination of the
line-center wavelength. We use the calculations of
\citet{1999A&AS..135..359C}, for densities $n \le 10^4~\pcc{}$ and
interpolated at a temperature of 9100~K\@. The resulting fine
structure contribution to the velocity dispersion,
$\sigma^2_\textrm{fs}$, and rest wavelengths, $\lambda_0$, are given
in the first two columns of Table~\ref{tab:fine}.

\begin{table*}
  \centering
  \caption{Rest wavelengths and RMS widths of Balmer lines}
  \label{tab:fine}
  \newcommand\C[1]{\multicolumn{1}{c}{#1}}
  \setlength\tabcolsep{1.5\tabcolsep}
  \begin{tabular}{l rr c rr}\toprule
    & \C{$\sigma^2_\textrm{fs}$} & \C{$\lambda_0$} 
    &\quad & \C{$V(\mathrm{Baldwin})$} & \C{$V(\mathrm{corrected})$} \\
    line & \C{(\kmssq{})} & \C{(\AA{})}&   &\C{(\kms{})} &\C{(\kms{})} \\
    \midrule
    H$\alpha$   & 10.233 & 6562.7910   & & $-0.7$ & 0.62      \\ 
    H$\beta$    & 5.767  & 4861.3201   & & 0.2 & 0.81      \\
    H$\gamma$   & 4.635  & 4340.4590   & & 1.3 & 0.68      \\
    \bottomrule
  \end{tabular}
\end{table*}

Many previous studies have used outdated and incorrect values for the
rest wavelengths, with a value of 6562.82~\AA{} being commonly assumed
for \Ha{}, rather than the correct value of 6562.7910~\AA{}\@.  This
leads to systematic errors of order 2~\kms{} in the derived radial
velocities of the Balmer lines, as can be seen, for example, in Fig.~6
of \citet{2000ApJS..129..229B}. However, as is shown in
Table~\ref{tab:fine}, these discrepancies disappear when the correct
rest wavelengths are used.

\end{appendices}
\bibliography{All-Sorted,Raga,Extras,ads-orion}

\end{document}